\documentclass[apj,tighten,twocolappendix]{emulateapj}

\usepackage{xspace}

\newcommand{\kms}{km s$^{-1}$\xspace}
\newcommand{\HI}{{\rm H\,{\scriptsize I}}\xspace}
\newcommand{\HII}{{\rm H\,{\scriptsize II}}\xspace}

\begin{document}
\slugcomment{Corrected: June 19, 2013}

\title{\HI Absorption Toward \HII Regions at Small Galactic Longitudes}

\author{C. Jones \& J. M. Dickey }
\affil{School of Mathematics and Physics, Private Bag 37, University of Tasmania, Hobart, 7001, Australia}

\author{J. R. Dawson}
\affil{School of Mathematics and Physics, Private Bag 37, University of Tasmania, Hobart, 7001, Australia; CSIRO Astronomy and Space Science, ATNF, PO Box 76, Epping, NSW, 1710, Australia}

\author{N. M. McClure-Griffiths}
\affil{CSIRO Astronomy and Space Science, ATNF, PO Box 76, Epping, NSW, 1710, Australia}

\author{L. D. Anderson}
\affil{Department of Physics, West Virginia University, Morgantown, WV, 26506, USA}

\and

\author{T. M. Bania}
\affil{Institute for Astrophysical Research, Department of Astronomy, Boston University, 725 Commonwealth Ave., Boston, MA, 02215, USA}

\begin{abstract}
We make a comprehensive study of \HI absorption toward \HII regions located within $|l|<10^{o}$.  Structures in the extreme inner Galaxy are traced using the longitude-velocity space distribution of this absorption.  We find significant \HI absorption associated with the Near and Far 3kpc Arms, the Connecting Arm, Bania's Clump 1 and the \HI Tilted Disk.  We also constrain the line of sight distances to \HII regions, by using \HI absorption spectra together with the \HII region velocities measured by radio recombination lines.
\end{abstract}


\section{Introduction}
The Extreme Inner Galaxy (EIG) has long been the subject of intense astrophysical study as it provides excellent opportunities to explore dynamics, phenomena (from stellar to galactic scales) and physical environments which do not exist in the large-scale Galactic disk.

Throughout this paper, we refer to the area inside of, and including, the 3kpc Arms as the EIG (i.e. $R_{Gal}\lesssim 4 \text{kpc}$). `Inner Galaxy' is a term already used to describe the areas of the Milky Way inside the Solar Circle, likewise the term `Galactic Center' (GC) usually refers to the relatively small area with a Galactocentric radius less than a few hundred parsecs.

Useful reviews of the EIG environment include \citet{MorrisSerabyn96} and \citet{Blitz93}, who discuss the interstellar medium (ISM) and structural components respectively.

Radio observations of the EIG region have been performed since the 1950s \citep[using the Dwingeloo 26 m antenna,][]{vanWoerden57}. These early studies discovered large-scale \HI features with non-circular motions \citep{Oort77}, and concentrated on understanding these individual structures, or particular objects.

The EIG has been extensively observed in CO. Molecular tracers probe denser material than neutral hydrogen (\HI) and CO is readily observed, therefore CO observations allow for analysis of regions in which the ISM is concentrated into structures such as arms and bars \citep{DameCO}. In contrast, observations of atomic gas trace diffuse interstellar clouds.

While \HI in the EIG has been well studied at low resolutions, it is only recently that high-resolution \HI data which cover the entire EIG region have become available \citep[i.e. ATCA \HI Galactic Center Survey (HIGCS)][]{HIGC}. These high-resolution \HI data allow an analysis of the beginnings of the spiral arms; the transition between orbits associated with the bar; a comparison to high-resolution molecular observations, dynamical models and molecular transitions; as well as investigations into the association of \HI with the Galactic wind \citep{HIGC}.

As a result of the variation in the temperature of interstellar hydrogen, \HI emission and absorption spectra probe different phases of the ISM. In most emission spectra it is the warmer components that dominate. However,  cool gas is readily observed in absorption against background continuum sources, where it may be disentangled from warmer material along the line of sight. One advantage to studying HI absorption in the EIG is that it probes this predominantly cool material, which tends to be more localised in space, and more closely confined to structural entities such as arms.

Previous \HI absorption studies have been vital to our understanding of the structure, rotation and the nature of atomic gas in the EIG region.  These include observations of absorption features associated with non-circular velocities, Radio Arc non-thermal filaments as well as particular objects including SgrA* \citep[][and references therein]{Lang10}.

While high-resolution \HI absorption measurements have been made towards several bright, or otherwise interesting, EIG continuum sources \citep[][and references therein]{Uchida92, Roy03, Lang10} a complete \HI absorption study of the EIG region has not been attempted. This present \HI absorption survey constitutes the most complete study of \HI absorption against the continuum emission from the entire sample of \HII regions known with $|l|<10^{o}$.  This study is only possible due to recent \HII region discovery surveys (which provide a list of target continuum regions with which to measure absorption against) and improved resolution in \HI surveys that include the GC region.

In addition to providing a sample of bright continuum sources against which to measure \HI absorption, \HII regions also provide an important secondary tracer of Galactic structure: the \HII regions themselves. Galactic \HII regions are the formation sites of massive OB stars, which have a main sequence lifetime of $\sim$tens of millions of years. As a result, Galactic \HII regions reveal the locations of current massive star formation, indicate the present state of the ISM, provide a unique probe of Galactic chemical evolution and are the archetypical tracers of Galactic spiral structure \citep{GBTHRDS}.

In this work we measure HI absorption against only those HII regions with known radio recombination line (RRL) velocities. This sub-sample is discussed in Section \ref{data}, and the method of HI absorption is described in Section \ref{HIabs}.

We then summarise the known EIG structures (Section \ref{GC}) and their locations in Longitude-Velocity ($lv$) space.  We plot these structures on an `$lv$ crayon diagram', and use the diagram to consider the EIG $lv$ distribution of \HI absorption, in Section \ref{HIdist}, and later for \HII regions (Section \ref{HIIdist}).

We combine the results from Sections \ref{HIdist} and \ref{HIIdist} to explore the Galactic distribution of \HII regions (Section \ref{Distances}) - through determining the lower limit of the line of sight distance to each \HII region based on its \HI absorption profile and systemic velocity.

Finally, a discussion of individual sources appears as Appendix \ref{individual}.

\section{Data \& Source Selection }
\label{data}
Large scale, high resolution astronomical surveys are now publicly available in many wavelength regimes.  This work uses large-scale \HI data and radio continuum maps.

\subsection{Radio Continuum }
\label{contin}
Radio continuum maps were sourced from the NRAO VLA Sky Survey \citep[NVSS]{NVSS} and the Southern Galactic Plane Survey  \citep[SGPS]{SGPS}. 

The NVSS covers 82\% of the sky (north of $\delta = -40^{o}$) at 1.4GHz, resulting in 2326 4x4 degree continuum cubes of Stokes parameters and a catalog of continuum emission sources.  Only the Stokes I maps were used for this work.  It should be noted that the NVSS maps do not include zero spacing (u,v) information and therefore many, larger, diffuse emission regions, particularly those in the \citet{LPH96} catalog, are not detected.  

\subsection{Neutral Hydrogen, \HI }
\label{HIdata}
 For this work, \HI absorption spectra were extracted from the two SGPS datasets ($5^{o}<|l|<10^{o}$) and the ATCA \HI Galactic Center Survey \citep[$5^{o}<|l|$, HIGCS][]{HIGC}.  Observations for the SGPS (I \& II) and ATCA HIGCS were performed with the Australia Telescope Compact Array (ATCA) and supplemented with data from the Parkes Radio Telescope.  The three surveys provide continuous coverage of the inner Galactic plane ($253^{o}<l<20^{o}$) at $\sim$2' resolution.

\subsection{Radio Recombination Lines}
 \label{RRLs}
Catalogues of RRLs provide systemic velocities for \HII regions.  Large-scale surveys of RRLs from \HII regions were performed during the 1960's to 1980's.  More recently, the Green Bank Telescope \HII Region Discovery Survey \citep[GBTHRDS]{GBTHRDS} covered $343^{o}<l<67^{o}$ and detected RRLs from 448 new \HII regions, effectively doubling the number known in that longitude range.  The GBTHRDS is complete to 180 mJy at 9 GHz, and is able to detect all \HII regions ionised by a single O-star to a distance of 12 kpc. 

In addition, the GBTHRDS also includes a catalog of known \HII regions as of 2010.  For the $|l|<10^{o}$ region, this includes the combined works of \citet{Downes80}, \citet{WAM82}, \citet{CH87}, \citet{Lockman89}, \citet{LPH96} and \citet{Sewilo04}.  The GBTHRDS team carefully compiled this ``known'' catalog, removing duplicate sources through radio continuum and mid-infrared inspection.  However they note that it is ``likely to contain some residual contamination and duplicate entries''.  The combination of this ``known'' catalog of \HII regions and the GBTHRDS source list, within $|l|<10^{o}$, provided the sample list of regions for this work.  Both the GBTHRDS catalog and the compilation of previous catalogs can be found at \url[http://www.cv.nrao.edu/hrds/]{http://www.cv.nrao.edu/hrds/}. 

\subsubsection{\HII Regions Selected} 
There are nearly 200 known \HII regions in the range $|l|<10^{o}\text{, }|b|<1.5^{o}$ with observed RRL velocities.  \HI absorption spectra were extracted towards a total of 151 of these \HII regions (see Figure Set \ref{Fig1}).  The remaining \HII regions were either not visible in the NVSS continuum maps (also used by the GBTHRDS), usually diffuse \HII regions from the \citet{LPH96} catalog, or \HII regions with coordinates that could refer to several continuum sources - see Table \ref{Table1}.  Therefore this study obtains \HI absorption spectra towards over 80\% of known \HII regions with $|l|<10^{o}$.  The `name' for each \HII region is taken from the RRL catalog from which it was sourced.

\begin{table}[h!]
\centering
\begin{tabular}{lll}
\hline
\hline
\HII Region	&	Reference	&	Note\\
\hline
G351.265$+$01.019	&	GBTHRDS (2011)		&	NV\\	
G351.590$+$00.183	&	\citet{Lockman89}	&	MS\\		
G353.035$+$00.748	&	\citet{Lockman89}	&	MS\\		
G353.083$+$00.358	&	\citet{Lockman89}	&	MS\\		
G357.998$-$00.159	&	\citet{Lockman89}	&	DC\\		
G358.319$-$00.414	&	\citet{LPH96}		&	NV\\	
G358.623$-$00.066	&	\citet{CH87}			&	DC\\
G358.661$-$00.575	&	\citet{LPH96}		&	NV\\	
G358.664$-$00.575	&	\citet{LPH96}		&	NV\\	
G358.974$-$00.021	&	\citet{LPH96}		&	NV\\	
G359.186$-$00.026	&	\citet{CH87}			&	DC\\
G359.730$-$00.407	&	\citet{Downes80}		&	NV\\	
G359.783$+$00.040	&	GBTHRDS (2011)		&	NV\\	
G359.929$+$00.045	&	GBTHRDS (2011)		&	NV\\	
G000.394$-$00.540	&	\citet{Downes80}		&	NV\\	
G000.521$+$00.178	&	\citet{LPH96}		&	NV\\	
G000.605$+$00.325	&	\citet{LPH96}		&	NV\\	
G000.656$-$00.058	&	\citet{Downes80}		&	NV\\	
G000.829$+$00.193	&	\citet{Downes80}		&	NV\\	
G001.323$+$00.086	&	\citet{CH87}			&	DC\\
G002.303$+$00.243	&	\citet{Lockman89}	&	MS\\		
G005.049$+$00.254	&	\citet{Lockman89}	&	NV\\		
G005.332$+$00.081	&	\citet{LPH96}		&	MS\\	
G006.616$-$00.545	&	\citet{LPH96}		&	NV\\	
G006.667$-$00.247	&	\citet{Lockman89}	&	NV\\		
G006.979$-$00.250	&	\citet{Lockman89}	&	NV\\		
G007.002$-$00.015	&	\citet{LPH96}		&	NV\\	
G007.299$-$00.116	&	\citet{Lockman89}	&	NV\\		
G007.387$+$00.668	&	\citet{Lockman89}	&	NV\\		
G008.415$+$00.033	&	\citet{LPH96}		&	NV\\	
G008.786$-$00.034	&	\citet{LPH96}		&	NV\\	
G009.176$+$00.032	&	\citet{LPH96}		&	DC\\	

\hline 
\hline
\end{tabular}
\caption{\HII regions that were \textit{not} included in this work.  Notes -  NV: not visible at the SGPS pixel scale; MS: many continuum sources present at this location; DC: duplicate (in both catalogs).} 
\label{Table1}
\end{table}

\section{Extraction of the \HI Absorption Spectra}
\label{HIabs}
The hyperfine transition that creates the 21-cm \HI line is often seen in both emission and absorption from the same region - indeed for most continuum sources a mixture of emission and absorption is observed.  Therefore a method is required to separate the two.

\subsection{Emission/Absorption Method}
\label{EAmethod}
The emission/absorption method \citep[described in detail by][]{Kolpak03} compares foreground cloud absorption with continuum emission from a background target.  Absorption, $e^{-\tau}$, is derived by comparing the brightness temperature as a function of velocity ($v$) both on ($T_{on}$ ) and off (i.e., the emission spectrum, $T_{off}$) the continuum source.  Continuum maps were inspected with the KARMA package \citep{KARMA} to ascertain the pixel positions for `on' and `off' spectra to be extracted from the \HI cubes; one `on' and three `off' source positions were chosen in accordance with the criteria identified in \citet{Jones12}.

The simplest radiative transfer situation gives:

\begin{equation}
T_{on}(v)=(T_{bg}+T_{cont})e^{-\tau}(v)+T_{s}(v)(1-e^{-\tau(v)})
\end{equation}

\noindent where $T_{cont}$ is the continuum source brightness temperature, $T_{s}$ is the spin temperature of the foreground cloud, and $T_{bg}$ represents any other background contribution. Assuming that both the “on” and “off” spectra sample the same gas, subtraction of one from the other removes the common $T_{s}(v)[1-e^{-\tau}(v)]$ and $T_{bg}$ terms allowing optical depth to be measured directly. The absorption is then
given by:

\begin{equation}
e^{-\tau}=\frac{T_{on}-T_{off}}{T_{cont}}
\label{AbsEq}
\end{equation}

The quality of an absorption spectrum is not determined by radiometer noise, but rather the accuracy of estimating the emission both on and around the background continuum source.  As such, we require absorption to be significant in relation to both emission fluctuations and the noise in the baseline of the absorption spectra (in a region without absorption features).

For each absorption spectrum, velocity channels with significant absorption were selected for analysis.  Significant absorption is defined to satisfy both:
\begin{itemize}
\item significance at the $3\sigma_{e^{-\tau}}$ level, where $\sigma_{e^{-\tau}}$ is calculated from the emission fluctuation envelope (the difference in emission between `off' sources).
\item significance at the $3\sigma_{rms_{e^{-\tau}}}$ level, where $\sigma_{rms_{e^{-\tau}}}$ is the fluctuation in the baseline of the absorption spectrum.
\end{itemize}

The NVSS continuum maps are biased towards smaller continuum temperatures (see \S \ref{contin}) as they do not include all diffuse continuum emission.  However, as $T_{cont}$ acts as a scaling factor for $e^{-\tau}$ (see Equation \ref{AbsEq}), $\sigma_{e^{-\tau}}$ and $\sigma_{rms_{e^{-\tau}}}$ will also scale proportionately with any change in continuum temperature.  

\begin{figure}
\includegraphics[width=\columnwidth]{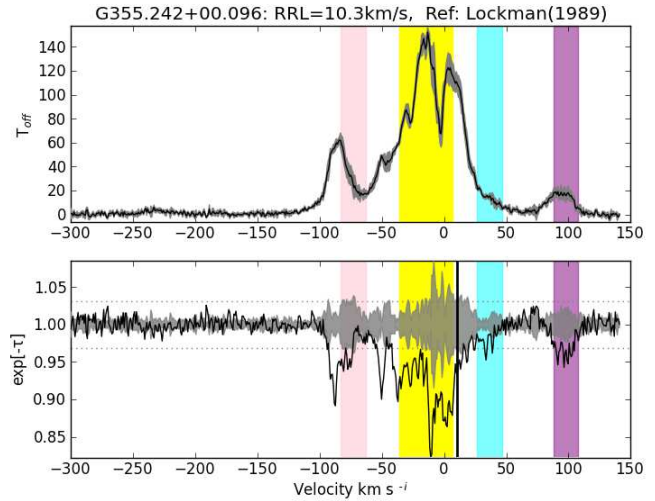}
\caption{Figure Set: \HI emission/absorption spectrum pairs.  In each figure, the top panel shows the emission spectra.  The emission is shown by the solid line (this is the average of the three `off' positions, see \ref{EAmethod}) and the emission envelope (difference between the `off' positions) is shown in grey. Absorption, $e^{-\tau}$, is displayed in the bottom panel.  The \HI absorption spectrum (see \ref{AbsEq}) is shown by the solid line and the grey envelope signifies $3\sigma_{e^{-\tau}}$ (calculated from the emission envelope, see \S \ref{HIabs}). The absorption panel also displays the RRL velocities of the \HII region (solid vertical lines) and the fluctuation in the baseline of the absorption spectrum ($3\sigma_{rms_{e^{-\tau}}}$) (horizontal dotted lines).  The \HII region name, RRL velocity and reference are shown as well as the expected velocity ranges of EIG features (see \S \ref{GC}) with the same color system as Figure \ref{Fig2}.  \label{Fig1}}
\end{figure}

Emission and absorption spectrum pairs toward each \HII region appear in Figure Set \ref{Fig1}.

\section{Longitude-Velocity Overview of the Extreme Inner Galaxy }
\label{cayondiagramsection}
Absorption spectra along lines of sight through the Galactic disk within the longitude range $|l|<10^{o}$ are complex and difficult to interpret.  This longitude region includes structures associated with the GC and EIG ($R_{Gal}\lesssim4\text{kpc}$), with highly non-circular motions; as well as features with velocities consistent with circular disk rotation ($R_{Gal}\gtrsim4\text{kpc}$).

\begin{figure}[h!]
\centering
\includegraphics[width=\columnwidth]{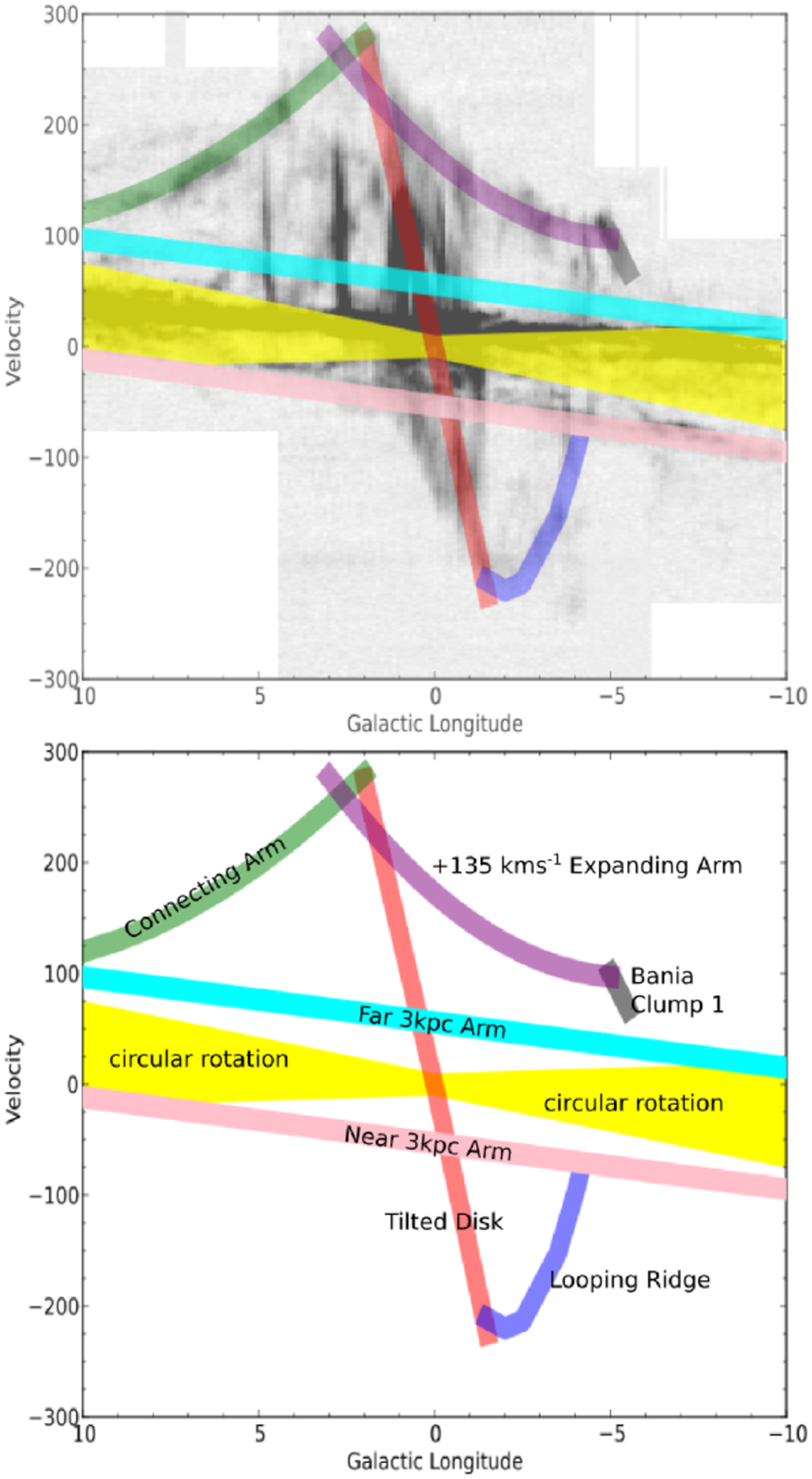}
\caption{Longitude-velocity ``crayon'' diagram for $l<|10^{o}|$, $b<|1.5|^{o}$.  Top panel - the `crayon' features overlaid on CO emission map \citep{DameCO}.  Bottom panel - the `crayon' features (each with a velocity width of 20 \kms).  The `crayon' color system is as follows: yellow - circular rotation allowed velocity envelope ; green - Connecting Arm; purple - +135kms$^{-1}$ Expanding Arm; grey - Bania's Clump 1; red - Tilted Disk; cyan - Far 3kpc Arm; yellow - velocities allowed by circular disk rotation; pink - Near 3kpc Arm; blue - Looping Ridge. While not explicitly labeled in the crayon diagram, Bania's Clump 2 can be seen as the thick vertical CO feature at $l\sim3^{o}$, $0\lesssim v \lesssim 200$ \kms in the top panel. (A color version of this figure is available in the online journal.) \label{Fig2}}
\end{figure}

Here we use the summary of EIG structures (\S \ref{GC}) to construct an `$lv$ crayon plot' (Figure \ref{Fig2}), marking prominent EIG features based on the integrated intensity of $^{12}$CO in the same $l,b$ range \citep[from][]{DameCO}.  

CO traces denser material than \HI and therefore picks out the densest features.  In the inner Galaxy, atomic gas often acts to shield associated regions of molecular gas from photodissociation \citep{DickeyLockman90}.  Therefore \HI absorption features may be identified with known EIG molecular emission features using correlations in velocity structure \citep{Lang10}.

As a result, this plot provides a useful reference, which we use to consider the $lv$ distribution of \HI absorption, \S \ref{HIdist}, (and later \HII region RRL velocities, \S \ref{HIIdist}).  
 
\subsection{Structures in the Extreme Inner Galaxy}
\label{GC}
Structures in the EIG include a long, thin bar, a shorter, boxy-bulge bar, the Near and Far 3 kpc arms, tilted \HI inner disk or ring, central molecular zone, and thin twisted 100 pc ring \citep{HIGC}.   In addition to these more prominent structures, recent $lv$ diagrams from \HI and CO observations show many `clumpy' sub-structures, not seen in previous EIG models \citep{BSW10}.
  
The angular extent of some of these EIG features is quite large: the Near 3kpc Arm is observed to $l<348^{o}$, and to surround all \HI emission associated with the EIG region, a latitude range of at least $|b|\leq8^{o}$ is required \citep{BurtonLiszt83} - well beyond the range of known \HII regions ($|b|<\sim2^{o}$).

Many of these features are not often explicitly discussed in the literature and precise distances are usually unknown \citep{Fux99}.  A summary of the EIG gas structures, many of which are visible in \HI absorption spectra appears below.  Often these objects have several names in the literature, or several distinct features have been given the same name by different authors. 

For a discussion of the evolution of the understanding of \HI and CO $lv$ models in the EIG see \citet{BSW10}.  \citet{BurtonLiszt83} provide a series of $lv$ diagrams with prominent features identified. 

\subsubsection{Near and Far 3kpc Arms}
\label{3kpcArms}
\paragraph{Near 3kpc Arm}
The Near 3kpc Arm or Expanding 3kpc Arm was discovered in the late 1950's and is known to lie in front of the GC \citep{vanWoerden57}.  However, it is not yet agreed whether the Near 3kpc Arm is a lateral arm surrounding the bar, or a small arm extending from the end of the bar, or an arm located where the bar meets its co-rotation radius \citep{RF11}.  Stretching over $35^{o}$ in longitude, the Near 3kpc Arm exists at `forbidden' velocities and its discovery provided vital early support for a Galactic bar \citep[][and references therein]{DameThaddeus08}.  The Near 3kpc Arm appears as the pink line in Figure \ref{Fig2}.

\paragraph{Far 3kpc Arm}
Despite the tendency for major anomalous velocity features in the GC to occur in positive and negative velocity pairs \citep{BurtonLiszt83}, it was originally thought that there was no far side counterpart to the Near 3kpc Arm \citep[][and references therin]{DameThaddeus08}.  \citet{Fux99} supposed the 135 \kms Arm (discussed below) was the feature symmetric to the Near 3kpc Arm - however Fux also noted compositional differences between the $+135$ \kms and Near 3kpc Arm, attributing these to an asymmetric spiral structure.  \citet{DameThaddeus08} reported the detection (in CO and then followed up in \HI) of the far side counterpart to the Near 3kpc Arm, named the Far 3kpc Arm.  The Far 3kpc Arm appears as the cyan line in Figure \ref{Fig2}.

\subsubsection{\HI Tilted/Nuclear Disk}
 The \HI inner tilted disk, proposed by \citet{LisztBurton80}, was the result of a full 3D analysis of all known \HI emission in the inner kiloparsec of the Galaxy.  It was modeled by a series of closed elliptical gas orbits.  The disk is oriented at $23.7^{o}$ with respect to the Galactic plane and accounts for positive velocity \HI emission at $b<0^{o}, l>0^{o}$ and negative velocity gas at $b>0^{o}, l<0^{o}$ (HIGCS).  In Figure \ref{Fig2}, the Tilted Disk appears as the red line crossing through $(l,v)=(0,0)$.
 
\subsubsection{The Expanding Arm(s)}
\paragraph{+135kms$^{-1}$ Arm} 
The location of the +135km s$^{-1}$ Arm, or Expanding Arm, is contentious throughout the literature: \citet{Fux99} assumes it is the far side counterpart to the Near 3kpc Arm (see \S \ref{3kpcArms}), \citet{Bania80} argues that the 3kpc and +135kms$^{-1}$ Arms can not be described together as a kinematic ring, and \citet{BSW10} model the +135kms$^{-1}$ Arm as part of the end of the bar on the far side.  

The +135kms$^{-1}$ Arm is more clumpy than the Near 3kpc Arm \citep{Fux99} and extends nearly 30$^{o}$ in longitude and spans 3$^{o}$ in latitude near the GC ($-1^{o}<b<2^{o}$ at $l=359^{o}$) \citep{Uchida92}.  Distance estimates for the +135kms$^{-1}$ Arm vary; \citet{SM73} and \citet{Bania80} give galactocentric radii only (3.4 kpc and 2.8-3.5kpc respectively), whereas \citet{Uchida92} give a distance estimate of about 2kpc \textit{behind} the GC (i.e. $D_{los}>10$kpc).

In Figure \ref{Fig2}, the +135\kms Expanding Arm appears as the purple curve.

\paragraph{Bania's Clumps} The individual emission clumps that comprise the +135kms$^{-1}$ Arm probably either include Bania's Clumps 1 and 2 \citep{Bania80, Bania86}, or the two molecular cloud complexes are entering the dust lane shock \citep{Liszt08}.  A detailed discussion of the \HI properties of Bania's Clump 2 can be found in \citet{HIGC}.

Bania's Clump 1 is seen as the grey line in Figure \ref{Fig2}, whereas Bania's Clump 2 is seen as the thick vertical CO feature at $l \approx 3^{o}$, $\sim0<v<\sim200$ \kms in the CO emission map (top panel of Figure \ref{Fig2}).

\paragraph{$-135$\kms Feature}
Just as the Near 3kpc Arm has a nearly symmetrical velocity and spatial counterpart a $-135$\kms Feature is thought to be located in the foreground of the GC, but behind the Near 3kpc Arm, as it is seen in OH absorption \citep{Uchida92}. This feature is much less distinct than the $+135$\kms Arm, indeed \citet{Bania80} did not detect it.  This feature is not included in Figure \ref{Fig2}.

\subsubsection{Connecting Arm and Looping Ridge}
Two features - the Connecting Arm and Looping Ridge - are visible in CO and \HI emission, as well as in near infrared dust extinction (\citet{Marshall08} and HIGCS).  These features lead the bar major axis and are the location of strong shearing shocks, resulting in high velocities \citep{Fux99}.  

The extent of both structures in $l,b,v$ has been explored in detail by \citet{Marshall08} who use CO data to localise emission to specific $lv$ structures.

\paragraph{Connecting Arm - Positive Velocity Feature}
The Connecting Arm (at extreme positive velocities and longitudes), was named as it seems to link the nuclear ring/disk to the outer disk \citep{Fux99}.  The Arm was sufficiently prominent in \HI to be described as a distinct feature in early EIG surveys \citep{Liszt08}.  The location of the Connecting Arm, in front of or behind the GC, was originally unclear; it has been interpreted as part of the central bar on the far side of the GC, or as an artifact due to velocity crowding along the line of sight, but it is now accepted to be a near side dust lane \citep[][and references therein]{Fux99}.

The Connecting Arm appears as the green curve in Figure \ref{Fig2}.

\paragraph{Looping Ridge - Negative Velocity Feature}
The corresponding feature to the Connecting Arm (at negative velocities and longitudes) is not always treated as a distinct feature \citep{Liszt08} and remains unnamed, however \citet{HIGC} refer to the negative feature as the ``looping'' ridge.  \citet{Liszt08} suggests that the Looping Ridge may be (temporarily) starved of gas and hence more difficult to detect and analyse.

In Figure \ref{Fig2}, the Looping Ridge appears as the blue curve.

\section{Longitude-Velocity Distribution of \HI Absorption Toward the Extreme Inner Galaxy }
\label{HIdist}
\begin{figure}[h!]
\centering
\includegraphics[width=0.9\columnwidth]{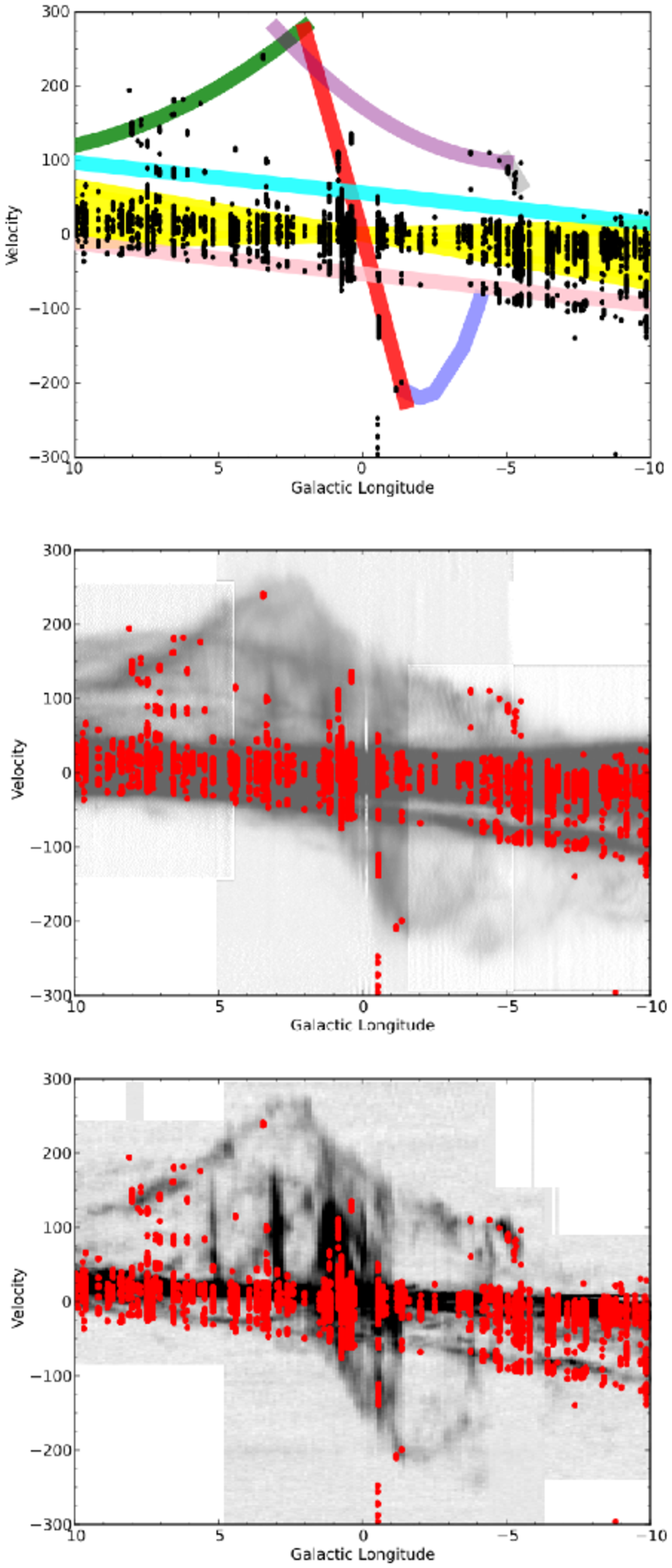}
\caption{$lv$ diagrams showing velocity channels with significant absorption for each \HII region overlaid onto: top panel - the `crayon' plot of EIG features (see Figure \ref{Fig2}); middle panel - \HI emission (the intensity map was created from the SGPS I \& II and ATCA HIGCS); bottom panel - CO emission (see Figure \ref{Fig2}). (A color version of this figure is available in the online journal.)  \label{Fig3}}
\end{figure}

Figure \ref{Fig3} displays the \HI absorption in $lv$ space, and compares this distribution with the EIG structures (\S \ref{GC}), \HI and CO emission.

Table \ref{Table2} notes if significant \HI absorption is associated with any EIG feature for each \HII region.

It is immediately obvious that the \HI absorption distribution is not random, but closely follows the identified EIG features.  This is not surprising as cold \HI gas, seen in absorption, is a good tracer of Galactic structure.

\HI absorption is associated with the allowed circular rotation velocities (as expected) as well as the Near and Far 3kpc Arms, Connecting Arm and Bania's Clump 1.

\begin{table*}[h!]
\tiny
\centering
\begin{tabular}{|lcccccc|lcccccc|}
\hline
\hline
Region & N3 & CA& TD &+E135 & BC1& F3&Region & N3 & CA& TD &E135 & BC1& F3\\
\hline
G350.004$+$00.438 & N & \nodata & \nodata & \nodata & \nodata & N & G000.284$-$00.478 & Y & \nodata & N & N & \nodata & N \\
G350.129$+$00.088 & Y & \nodata & \nodata & \nodata & \nodata & N & G000.361$-$00.780 & Y & \nodata & N & N & \nodata & Y \\
G350.177$+$00.017 & N & \nodata & \nodata & \nodata & \nodata & Y & G000.382$+$00.017 & Y & \nodata & N & Y & \nodata & N \\
G350.330$+$00.157 & Y & \nodata & \nodata & \nodata & \nodata & Y & G000.510$-$00.051 & Y & \nodata & N & N & \nodata & N \\
G350.335$+$00.107 & Y & \nodata & \nodata & \nodata & \nodata & N & G000.572$-$00.628 & Y & \nodata & N & N & \nodata & N \\
G350.524$+$00.960 & N & \nodata & \nodata & \nodata & \nodata & N & G000.640$+$00.623 & Y & \nodata & Y & N & \nodata & Y \\
G350.769$-$00.075 & N & \nodata & \nodata & \nodata & \nodata & N & G000.729$-$00.123 & Y & \nodata & N & N & \nodata & Y \\
G350.813$-$00.019 & N & \nodata & \nodata & \nodata & \nodata & Y & G000.838$+$00.189 & Y & \nodata & Y & Y & \nodata & Y \\
G350.996$-$00.557 & N & \nodata & \nodata & \nodata & \nodata & Y & G001.125$-$00.105 & Y & \nodata & N & Y & \nodata & Y \\
G351.028$+$00.155 & Y & \nodata & \nodata & \nodata & \nodata & Y & G001.149$-$00.062 & Y & \nodata & Y & N & \nodata & Y \\
G351.047$-$00.322 & N & \nodata & \nodata & \nodata & \nodata & N & G001.324$+$00.104 & N & \nodata & N & N & \nodata & N \\
G351.192$+$00.708 & N & \nodata & \nodata & \nodata & \nodata & N & G001.330$+$00.088 & Y & \nodata & N & N & \nodata & Y \\
G351.201$+$00.483 & N & \nodata & \nodata & \nodata & \nodata & N & G001.488$-$00.199 & Y & \nodata & N & Y & \nodata & Y \\
G351.358$+$00.666 & N & \nodata & \nodata & \nodata & \nodata & Y & G002.009$-$00.680 & Y & N & \nodata & N & \nodata & N \\
G351.359$+$01.014 & N & \nodata & \nodata & \nodata & \nodata & N & G002.404$+$00.068 & N & N & \nodata & N & \nodata & N \\
G351.467$-$00.462 & N & \nodata & \nodata & \nodata & \nodata & N & G002.418$-$00.611 & N & N & \nodata & N & \nodata & N \\
G351.601$-$00.348 & N & \nodata & \nodata & \nodata & \nodata & Y & G002.510$-$00.028 & Y & N & \nodata & Y & \nodata & N \\
G351.662$+$00.518 & Y & \nodata & \nodata & \nodata & \nodata & N & G002.611$+$00.135 & Y & N & \nodata & N & \nodata & N \\
G351.691$+$00.669 & N & \nodata & \nodata & \nodata & \nodata & N & G002.819$-$00.132 & N & N & \nodata & N & \nodata & N \\
G352.233$-$00.151 & Y & \nodata & \nodata & \nodata & \nodata & N & G002.901$-$00.006 & Y & N & \nodata & N & \nodata & N \\
G352.313$-$00.440 & Y & \nodata & \nodata & \nodata & \nodata & Y & G002.961$-$00.053 & Y & N & \nodata & N & \nodata & Y \\
G352.398$-$00.057 & Y & \nodata & \nodata & \nodata & \nodata & N & G003.270$-$00.101 & Y & N & \nodata & \nodata & \nodata & Y \\
G352.521$-$00.144 & Y & \nodata & \nodata & \nodata & \nodata & N & G003.342$-$00.079 & Y & Y & \nodata & \nodata & \nodata & Y \\
G352.610$+$00.177 & N & \nodata & \nodata & \nodata & \nodata & N & G003.439$-$00.349 & Y & N & \nodata & \nodata & \nodata & N \\
G352.611$-$00.172 & Y & \nodata & \nodata & \nodata & \nodata & N & G003.449$-$00.647 & Y & Y & \nodata & \nodata & \nodata & N \\
G352.866$-$00.199 & Y & \nodata & \nodata & \nodata & \nodata & N & G003.655$-$00.111 & Y & N & \nodata & \nodata & \nodata & N \\
G353.186$+$00.887 & N & \nodata & \nodata & \nodata & \nodata & N & G003.928$-$00.116 & Y & N & \nodata & \nodata & \nodata & Y \\
G353.218$-$00.249 & Y & \nodata & \nodata & \nodata & \nodata & N & G003.949$-$00.100 & Y & N & \nodata & \nodata & \nodata & N \\
G353.381$-$00.114 & Y & \nodata & \nodata & \nodata & \nodata & N & G004.346$+$00.115 & N & N & \nodata & \nodata & \nodata & N \\
G353.398$-$00.391 & N & \nodata & \nodata & \nodata & \nodata & N & G004.412$+$00.118 & Y & N & \nodata & \nodata & \nodata & Y \\
G353.557$-$00.014 & Y & \nodata & \nodata & \nodata & \nodata & Y & G004.527$-$00.136 & Y & N & \nodata & \nodata & \nodata & Y \\
G354.200$-$00.054 & Y & \nodata & \nodata & \nodata & \nodata & N & G004.568$-$00.118 & Y & N & \nodata & \nodata & \nodata & N \\
G354.418$+$00.036 & N & \nodata & \nodata & \nodata & \nodata & N & G005.193$-$00.284 & Y & Y & \nodata & \nodata & \nodata & N \\
G354.486$+$00.085 & Y & \nodata & \nodata & \nodata & \nodata & Y & G005.479$-$00.241 & Y & N & \nodata & \nodata & \nodata & Y \\
G354.588$+$00.007 & Y & \nodata & \nodata & \nodata & N & N & G005.524$+$00.033 & Y & Y & \nodata & \nodata & \nodata & Y \\
G354.610$+$00.484 & Y & \nodata & \nodata & \nodata & N & Y & G005.633$+$00.240 & N & Y & \nodata & \nodata & \nodata & N \\
G354.664$+$00.470 & N & \nodata & \nodata & \nodata & N & N & G005.899$-$00.427 & N & Y & \nodata & \nodata & \nodata & N \\
G354.665$+$00.247 & N & \nodata & \nodata & \nodata & Y & N & G006.014$-$00.364 & Y & N & \nodata & \nodata & \nodata & N \\
G354.717$+$00.293 & N & \nodata & \nodata & \nodata & Y & N & G006.083$-$00.117 & Y & N & \nodata & \nodata & \nodata & Y \\
G354.934$+$00.327 & Y & \nodata & \nodata & \nodata & Y & Y & G006.148$-$00.635 & Y & N & \nodata & \nodata & \nodata & N \\
G354.979$-$00.528 & N & \nodata & \nodata & \nodata & N & N & G006.160$-$00.608 & Y & N & \nodata & \nodata & \nodata & N \\
G355.242$+$00.096 & Y & \nodata & \nodata & Y & \nodata & Y & G006.225$-$00.569 & N & Y & \nodata & \nodata & \nodata & N \\
G355.344$+$00.145 & Y & \nodata & \nodata & Y & \nodata & Y & G006.398$-$00.474 & N & Y & \nodata & \nodata & \nodata & N \\
G355.532$-$00.100 & Y & \nodata & \nodata & Y & \nodata & N & G006.553$-$00.095 & Y & Y & \nodata & \nodata & \nodata & Y \\
G355.581$+$00.288 & Y & \nodata & \nodata & Y & \nodata & Y & G006.565$-$00.297 & N & Y & \nodata & \nodata & \nodata & Y \\
G355.661$+$00.382 & Y & \nodata & \nodata & Y & \nodata & N & G007.041$+$00.176 & Y & N & \nodata & \nodata & \nodata & Y \\
G355.696$+$00.350 & Y & \nodata & \nodata & Y & \nodata & N & G007.176$+$00.087 & Y & N & \nodata & \nodata & \nodata & N \\
G355.700$-$00.100 & Y & \nodata & \nodata & N & \nodata & N & G007.254$-$00.073 & Y & N & \nodata & \nodata & \nodata & Y \\
G355.734$+$00.138 & Y & \nodata & \nodata & Y & \nodata & N & G007.266$+$00.183 & Y & Y & \nodata & \nodata & \nodata & N \\
G355.801$-$00.253 & Y & \nodata & \nodata & N & \nodata & N & G007.299$-$00.116 & N & N & \nodata & \nodata & \nodata & N \\
G356.230$+$00.066 & Y & \nodata & \nodata & N & \nodata & N & G007.420$+$00.366 & Y & N & \nodata & \nodata & \nodata & Y \\
G356.235$+$00.642 & Y & \nodata & \nodata & Y & \nodata & N & G007.466$-$00.279 & Y & N & \nodata & \nodata & \nodata & N \\
G356.307$-$00.210 & Y & \nodata & \nodata & N & \nodata & N & G007.472$+$00.060 & Y & N & \nodata & \nodata & \nodata & Y \\
G356.470$-$00.001 & Y & \nodata & \nodata & N & \nodata & Y & G007.700$-$00.079 & N & N & \nodata & \nodata & \nodata & N \\
G356.560$-$00.086 & Y & \nodata & \nodata & N & \nodata & N & G007.768$+$00.014 & Y & Y & \nodata & \nodata & \nodata & N \\
G356.650$+$00.129 & Y & \nodata & \nodata & Y & \nodata & N & G007.806$-$00.621 & Y & Y & \nodata & \nodata & \nodata & N \\
G357.484$-$00.036 & N & \nodata & \nodata & N & \nodata & Y & G008.005$-$00.484 & Y & Y & \nodata & \nodata & \nodata & N \\
G357.970$-$00.169 & Y & \nodata & \nodata & N & \nodata & N & G008.006$-$00.156 & Y & Y & \nodata & \nodata & \nodata & Y \\
G357.998$-$00.159 & Y & \nodata & \nodata & N & \nodata & N & G008.094$+$00.085 & N & N & \nodata & \nodata & \nodata & N \\
G358.319$-$00.414 & N & \nodata & \nodata & N & \nodata & N & G008.103$+$00.340 & Y & N & \nodata & \nodata & \nodata & N \\
G358.379$-$00.840 & N & \nodata & \nodata & N & \nodata & N & G008.137$+$00.228 & N & N & \nodata & \nodata & \nodata & N \\
G358.530$+$00.056 & N & \nodata & N & N & \nodata & N & G008.362$-$00.303 & Y & N & \nodata & \nodata & \nodata & N \\
G358.552$-$00.025 & N & \nodata & N & N & \nodata & N & G008.373$-$00.352 & Y & Y & \nodata & \nodata & \nodata & N \\
G358.616$-$00.076 & Y & \nodata & N & N & \nodata & N & G008.432$-$00.276 & Y & N & \nodata & \nodata & \nodata & Y \\
G358.623$-$00.066 & Y & \nodata & Y & N & \nodata & N & G008.666$-$00.351 & Y & N & \nodata & \nodata & \nodata & N \\
G358.633$+$00.062 & Y & \nodata & Y & N & \nodata & N & G008.830$-$00.715 & Y & N & \nodata & \nodata & \nodata & N \\
G358.652$-$00.078 & N & \nodata & N & N & \nodata & N & G008.865$-$00.323 & Y & N & \nodata & \nodata & \nodata & N \\
G358.680$-$00.087 & N & \nodata & Y & N & \nodata & N & G009.178$+$00.043 & Y & N & \nodata & \nodata & \nodata & N \\
G358.694$-$00.075 & N & \nodata & Y & N & \nodata & N & G009.615$+$00.198 & Y & N & \nodata & \nodata & \nodata & N \\
G358.720$+$00.011 & Y & \nodata & N & N & \nodata & N & G009.682$+$00.206 & N & N & \nodata & \nodata & \nodata & N \\
G358.797$+$00.058 & Y & \nodata & N & N & \nodata & N & G009.717$-$00.832 & Y & N & \nodata & \nodata & \nodata & N \\
G358.827$+$00.085 & N & \nodata & N & N & \nodata & N & G009.741$+$00.842 & Y & N & \nodata & \nodata & \nodata & N \\
G359.159$-$00.038 & Y & \nodata & N & N & \nodata & N & G009.875$-$00.749 & Y & Y & \nodata & \nodata & \nodata & N \\
G359.277$-$00.264 & N & \nodata & N & N & \nodata & N & G009.925$-$00.745 & Y & N & \nodata & \nodata & \nodata & N \\
G359.432$-$00.086 & Y & \nodata & N & N & \nodata & N & G009.982$-$00.752 & Y & N & \nodata & \nodata & \nodata & N \\
G359.467$-$00.172 & Y & \nodata & N & N & \nodata & Y & \nodata & \nodata & \nodata & \nodata & \nodata & \nodata & \nodata \\
\hline 
\hline
\end{tabular}
\caption{Presence of significant \HI absorption in EIG features for each \HII region.  EIG features are listed in line of sight order.  N3= Near 3kpc Arm, CA= Connecting Arm, TD= \HI Tilted Disk, E135= +135\kms Expanding Arm, BC1= Bania's Clump 1, F3= Far 3kpc Arm.} 
\label{Table2}
\end{table*}

\subsection{\HI Absorption Associated with the 3kpc Arms}
\label{HIin3kpcArms}
The CO emission from both the Near and Far 3kpc Arms is contained within $|b|<1^{o}$ \citep{DameThaddeus08}, similar to the Galactic latitude range of the \HII regions in this work ($|b|\lesssim1.5^{o}$).  Furthermore, both 3kpc Arms are thought to span $|l|\lesssim13^{o}$, which includes the whole longitude range of this work.  Therefore, if an \HII region is located behind either of the 3kpc Arms, \HI absorption should be visible at velocities pertaining to that arm.  

Figure \ref{Fig3} demonstrates that significant \HI absorption is seen toward the Near 3kpc Arm at all longitudes; although there is a conspicuous gap in absorption at longitudes $\sim356<l^{o}<\sim358$, consistent with a paucity of \HII regions for which to measure absorption towards.  Indeed 67$\%$ \HII regions display absorption associated with the Near 3kpc Arm.

There is less absorption associated with the Far 3kpc Arm than with the Near ($\sim1.0:3.3$), with the site of greatest absorption for the Far 3kpc Arm centered at $l\approx7^{o}$ (see Figure \ref{Fig4}).  The disparity in the amount of \HI absorption may be an effect of the smaller latitude extent of the Far 3kpc Arm, which is particularly thin in the fourth quadrant \citep{DameThaddeus08}.

Both the HIGCS and \citet{DameThaddeus08} report a bifurcation in the velocities Far 3kpc Arm (in $lv$ space) at $l<6^{o}$.  There is limited evidence of this bifurcation at longitudes extending to $\approx7^{o}$, the best example of this is in the absorption spectrum of G007.176+00.087 (see Figure Set \ref{Fig1}).  The Near 3kpc Arm also displays evidence of bifurcation, in both the \HI emission and absorption, near $l=358^{o}$ (see, for example, G358.616-00.076, G358.623-00.066, G358.633+00062 and G359.432-00.086 in Figure Set \ref{Fig1}).

\subsubsection{The Longitude-Velocity Location of the 3kpc Arms \label{3kpcLocation}}
The locus of each of the 3kpc Arms, as traced by \HI absorption, in $lv$ space was also investigated.  \citet{DameThaddeus08} provide $lv$ fits to the Near and Far 3kpc Arms and report a velocity dispersion of 21\kms for both Arms.  However they excluded large regions of longitude, within $|l|<10^{o}$, from the computation of the physical properties of each arm.

\begin{figure}[h!]
\centering
\includegraphics[width=\columnwidth]{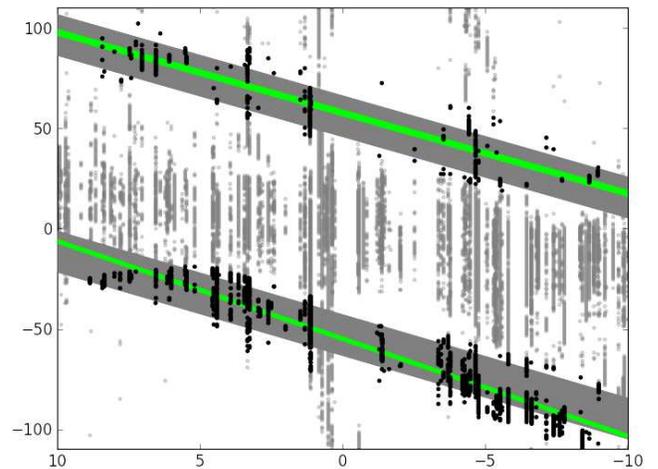}
\caption{Velocity channels of significant \HI absorption: those included in the tracing of the 3kpc Arms are shown in black, while those channels which were excluded from the analysis are in grey.  The fits of \citet{DameThaddeus08} are given by the grey bands (see Figure \ref{Fig2}), the fits from this analysis are displayed in green. (A color version of this figure is available in the online journal.)  \label{Fig4}}
\end{figure}

In order to investigate the locus of each Arm in $lv$ space, a subset of absorption channels were selected for analysis (see Figure \ref{Fig4}).  We included all channels within $1^{o}<|l|<9^{o}$ (between $9^{o}<|l|<10^{o}$ there is ambiguity between the 3kpc Arms and the circular rotation velocities and for $|l|<1^{o}$ there is ambiguity with the Tilted Disk), which had velocities outside the envelope of allowed circular-rotation velocities (i.e. $V_{circular-rotation}<|V|<110$\kms) and were not associated with either of Bania's Clumps.  Linear fits to these appear as Equations \ref{Nabs} and \ref{Fabs} for the Near and Far arms respectively.

Near 3kpc Arm:
\begin{eqnarray}
V_{N3kpc}=  -59.2+&4.12l \pm 8.67  \label{Nabs}\\
V_{F3kpc}=  +57.7+&4.02l  \pm 15.61 \label{Fabs} 
\end{eqnarray}

In both cases, the linear fits ($\pm 5\sigma$) of the structure as given by \HI absorption are consistent with the \citet{DameThaddeus08} fits from CO emission (see Figure \ref{Fig4}).

\section{Longitude-Velocity Distribution of \HII Regions Toward the Extreme Inner Galaxy }
\label{HIIdist}
\HII regions provide radio continuum sources to measure \HI absorption toward, but they also provide a secondary tracer of the EIG region - their own systemic velocities.

\begin{figure}[h]
\centering
\includegraphics[width=\columnwidth]{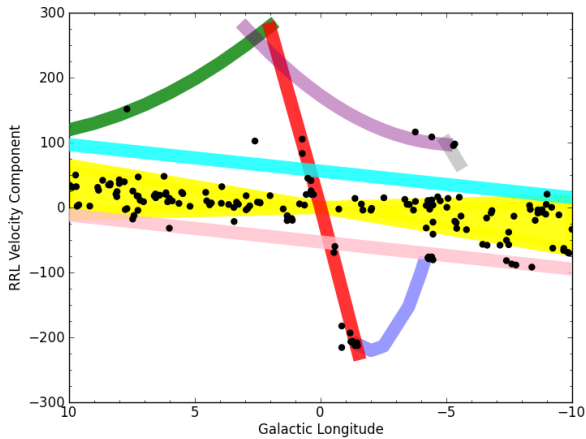}
\caption{Longitude-velocity `crayon' diagram (see Figure \ref{Fig2}) showing the distribution of \HII region RRL velocity components.(A color version of this figure is available in the online journal.)  \label{Fig5}}
\end{figure}

The $lv$ distribution of known \HII regions has previously been investigated by \citet{Anderson12}, however all \HII regions with highly non-circular motions (i.e. those of interest to this work) were excluded from their analysis.  The $lv$ distribution of \HII region RRL velocity components used in this work is shown in Figure \ref{Fig5} - note: for \HII regions with multiple velocity components, all are shown.  

Just as the distribution of \HI absorption was closely associated with known $lv$ features in the EIG region (\S \ref{HIdist}, Figure \ref{Fig3}), the systemic velocities of \HII regions also trace these structures.

The circular-rotation allowed velocities (yellow envelope in `crayon plots') account for $\sim85\%$ of the \HII region RRL velocity components. \citet{GreenMasers}, in a study of 6.7 GHz methanol masers near the EIG, find the same velocity range accounts for $\sim79\%$ of their sample.  However, only $\sim10\%$ of \HII regions with a single RRL velocity component are associated with EIG features.  A list of \HII regions with RRL velocities associated with an EIG structure appears in Table \ref{Table3}.

\subsection{\HII Regions Associated with $R_{Gal}\lesssim4$}
Until recently, it was believed that there are no known \HII regions inside of the 3kpc Arms, except in the Tilted Disk \citep{RF06}.  \citet{GreenMasers} found no significant 6.7 GHz methanol maser emission towards the +135km s$^{-1}$ Expanding Arm, nor the Connecting Arm; suggesting that the features are primarily gas that is not undergoing high-mass star formation.  This is in-keeping with observations of other early-type barred galaxies which show star formation in the central nuclear region and at the ends of the bar, but not in the dust lanes along the bar \citep{RFetal06}.  

Using the collated \HII region catalog of \citet{Pal03}, \citet{RFetal06} found no \HII regions associated with structures outside the Nuclear Disk within $|l|<2^{o}$.  The GBTHRDS recorded RRL velocity components from 21 previously unknown \HII regions within $|l|<2^{o}$, many (especially in Quadrant IV) with non-circular velocities - these \HII regions are included in the target list of this work.  However, as the \citet{RFetal06} study found, these new \HII regions are associated (in $lv$ space) with the Nuclear Disk and Looping Ridge intersection.  \citet{RFetal06} then investigated a wider longitude range, but could not identify any dust lane associated \HII regions.  It should be noted that \citet{RFetal06} did not rule out the possibility of undetected ultra-compact \HII regions in the dust lanes.

One diffuse \HII region, G007.700-0.079 identified by \citet{LPH96} (but not included in the \citet{Pal03} catalog used in the \citet{RFetal06} study) appears to have one of its RRL velocity components associated with the Connecting Arm.  There is also evidence of two 6.7 GHz methanol masers, tracers of current high-mass star formation, in the same part of $lv$ space \citep[see Figure 1 of][]{GreenMasers}.  In addition, there are four \HII region RRL velocity components associated with the +135kms$^{-1}$ Expanding Arm and/or Bania's Clump 1 (at $l,v=\sim-4^{o},\sim100$ \kms), as well as two 6.7 GHz methanol masers from the Methanol Multibeam survey \citep[cited in][]{GreenMasers}, however only two of these regions have single RRL velocity components (multi-RRL component \HII regions are probably the result of blending multiple emission sources along the line of sight).  Therefore there is evidence of some recent star formation in these structures.

The other \HII region of note is G002.611+0.135 as it is the only \HII region that distinctly lies outside the ``crayon'' lines that delineate EIG structures in Figure \ref{Fig3}. \citet{RFetal06} suggest that G002.611+0.135 could be associated with either their structure 'J' or \objectname[Bania's Clump 2]{Bania's Clump 2} \citep[see Figure 4 of][]{RFetal06}.  The latitude of the \HII region suggests a stronger association with Clump 2.  

\subsection{\HII Regions Associated with the 3kpc Arms}
Only recently has there been evidence of significant star formation \citep{Green09} and large numbers of \HII regions \citep{BaniaAnderson10} in the 3kpc Arms.  In emission from molecular clouds the signatures of the 3kpc Arms are clearly seen \citep{Bania80}, but the GBTHRDS was unable to discover many new \HII regions, in either of the Arms. However, both arms demonstrate high-mass star formation as traced by about fifty 6.7 GHz methanol masers \citep{Green09}.  

The certainty of associating \HII regions with the 3kpc Arms (in $lv$ space) is best in the longitude range of this study ($|l|<10^{o}$), as outside this limit the expected velocities of the 3kpc Arms overlap with circular-motion spiral arm models and the association becomes more ambiguous \citep{Green09}.

Inside $|l|<10^{o}$ there are eleven \HII region RRL velocity components consistent with the Near 3kpc Arm and two consistent with the Far 3kpc Arm.  This small number of RRL components does not allow for a repetition of the analysis of \S \ref{3kpcLocation} using \HII region RRL components rather than \HI absorption.

\section{Distance Constraints for \HII Regions from \HI Absorption }
\label{Distances}
The analysis of an \HI absorption spectrum towards a \HII region can constrain the line of sight distance to the \HII itself.

Due to the lack of a reliable rotation model for the inner $\sim$3kpc of the Milky Way, kinematic distances to objects near, or in, the EIG are the most difficult to ascertain.  However, it should be possible to provide distance constraints for \HII regions with allowed circular rotation systemic velocities, using \HI absorption associated with EIG features as approximate distance indicators.

In the Inner Galaxy, inside the Solar Circle, each velocity corresponds to two degenerate solutions for the kinematic distance - each equidistant from the tangent (subcentral) point.  This kinematic distance ambiguity can be resolved in cases where \HI absorption is present at the velocity of a known structure - which indicates the \HII region must be located behind the absorbing gas.

The distance arrangement of EIG features, listed in \S \ref{GC}, from the literature is assumed to be: Near 3kpc Arm, Connecting Arm, Tilted Disk, Looping Ridge, +135\kms Expanding Arm (and Bania's Clump 1), Far 3kpc Arm (however not all structures are expected along any single line of sight, see Figure \ref{Fig2}).

Therefore, if \HI absorption is seen at velocities corresponding to a particular feature, the \HII region must lie in, or beyond that structure.  In this way, we use \HI absorption as an indicator of the lower limit of the line of sight distance, $D_{los}$.  The RRL velocity of an \HII region also hints at its location, Table \ref{Table3} lists those \HII regions with systemic velocities beyond the range expected by circular rotation (see Figure \ref{Fig5}) and associated with EIG feature(s).

A discussion of each \HII region appears in Appendix \ref{individual}.

\begin{table}[h!]
\centering
\begin{tabular}{clccl}
\hline
\hline
\HII Region &\multicolumn{2}{c}{$V_{RRL}$} & Ref. & Association\\
\hline
\multicolumn{5}{c}{\textit{Single RRL Velocity Component \HII Regions}}\\
\hline
G$350.996-00.557$&\multicolumn{2}{c}{ $+20.5$}&	2	&Far 3kpc Arm\\
G$351.601-00.348$&\multicolumn{2}{c}{ $-91.8$}&3&Near 3kpc Arm\\
G$352.233-00.151$&\multicolumn{2}{c}{ $-88.6$}& 1 & Near 3kpc Arm\\
G$352.398-00.057$&\multicolumn{2}{c}{ $-87.0$}& 2& Near 3kpc Arm\\
G$352.611-00.172$&\multicolumn{2}{c}{ $-81.9$}& 2& Near 3kpc Arm\\
G$354.665+00.247$&\multicolumn{2}{c}{ $+97.8$}& 2& Bania's Clump 1?\\
G$354.717+00.293$&\multicolumn{2}{c}{ $+95.3$}& 1& Bania's Clump 1?\\
G$355.700-00.100$&\multicolumn{2}{c}{ $-76.1$}& 2 & Near 3kpc Arm\\
G$356.235+00.642$&\multicolumn{2}{c}{$+116.3$}& 2 & +135\kms Arm\\
G$358.530+00.056$&\multicolumn{2}{c}{$-212.6$}& 1 & Looping Ridge\\
G$358.552-00.025$&\multicolumn{2}{c}{$-208.5$}& 1 & Looping Ridge\\
G$358.616-00.076$&\multicolumn{2}{c}{$-212.6$}& 1 & Looping Ridge\\
G$358.623-00.066$&\multicolumn{2}{c}{$-212.0$}& 3 & Looping Ridge\\
G$358.652-00.078$&\multicolumn{2}{c}{$-211.2$}& 1 & Looping Ridge\\
G$358.680-00.087$&\multicolumn{2}{c}{$-208.3$}& 1 & Looping Ridge\\
G$358.694-00.075$&\multicolumn{2}{c}{$-207.8$}& 1 & Looping Ridge\\
G$358.720+00.011$&\multicolumn{2}{c}{$-206.1$}& 1 & Looping Ridge\\
G$358.797+00.058$&\multicolumn{2}{c}{$-206.6$}& 2 & Looping Ridge\\
G$358.827+00.085$&\multicolumn{2}{c}{$-193.3$}& 1 & Looping Ridge\\
G$359.432-00.086$&\multicolumn{2}{c}{$-60.0$}& 3 & Near 3kpc Arm\\
G$359.467-00.172$&\multicolumn{2}{c}{$-69.3$}& 1 & \textit{blended EIG features}\\
G$000.510-00.051$&\multicolumn{2}{c}{$+45.0$}& 4 & Far 3kpc Arm\\
G$003.439-00.349$&\multicolumn{2}{c}{$-21.6$}& 1 & Near 3kpc Arm?\\
G$007.472+00.060$&\multicolumn{2}{c}{$-17.8$}& 2 & Near 3kpc Arm\\
\hline
\multicolumn{5}{c}{\textit{Multiple RRL Velocity Component \HII Regions}}\\
\hline
G$355.532-00.100$&a&$+3.8$&1&\nodata\\
G$355.532-00.100$&b&$-22.5$&1&\nodata\\
G$355.532-00.100$&c&$-82.6$&1&Near 3kpc Arm\\
G$355.532-00.100$&d&$-41.1$&1&\nodata\\
G$355.581+00.028$&a&$108.7$&1&+135\kms Arm\\
G$355.581+00.028$&b&$-76.1$&1&Near 3kpc Arm\\
G$355.581+00.028$&c&$11.7$&1&\nodata\\
G$355.696+00.350$&a&$+3.0$&1&\nodata\\
G$355.696+00.350$&b&$-79.1$&1&Near 3kpc Arm\\
G$355.734+00.138$&a&$+10.7$&1&\nodata\\
G$355.734+00.138$&b&$-77.4$&1&Near 3kpc Arm\\
G$359.159-00.038$&a&$-182.5$&1& \textit{blended EIG features}\\
G$359.159-00.038$&b&$-215.6$&1& \textit{blended EIG features}\\
G$000.729-00.123$&a&$+105.3$&1&Tilted Disk\\
G$000.729-00.123$&b&$+83.2$&1&\nodata\\
G$006.014-00.364$&a&$+14.2$&1&\nodata\\
G$006.014-00.364$&b&$-31.9$&1&Near 3kpc Arm\\
G$007.700-00.079$& a& $-1.7$ & 5 & \nodata\\
G$007.700-00.079$&b&+151.7& 5&Connecting Arm\\
\hline 
\hline
\end{tabular}
\caption{\HII Regions with an RRL velocity associated (in $lv$ space) with known EIG structure(/s) - see Figure \ref{Fig5}. References for $V_{RRL}$ are as follows: 1 - GBTHRDS (2011), 2 - \citet{Lockman89}, 3 - \citet{CH87} and 4 - \citet{Downes80}, 5 - \citet{LPH96}.} 
\label{Table3}
\end{table}

\subsection{Kinematic Distances of Selected \HII Regions}
For \HII regions with systemic velocities associated with normal circular disk rotation (i.e. in the yellow envelope in Figure \ref{Fig5}, $R_{Gal}>4$kpc), a Kinematic Distance Ambiguity Resolution (KDAR) is attempted.  If a KDAR is achieved, the kinematic distance to the \HII region can then be calculated.

KDARs were achieved following these rules:
\begin{itemize}
\item if the \HII region RRL is consistent with normal circular disk rotation, and not with any EIG feature (see Table \ref{Table3}) (i.e. no kinematic distances are calculated for regions with $R_{Gal}<3$kpc).
\item FAR: If the \HI absorption spectrum displays sufficient absorption associated with EIG features, then the \HII region must be located at the `far' kinematic location - i.e. beyond the tangent point along the line of sight.  
\begin{itemize}
\item if $|V_{RRL}|<10$\kms a KDAR is attempted \textit{only} if the \HI absorption spectrum displays absorption associated with the Far 3kpc Arm, this is imposed in order to avoid confusion with EIG features at small velocities.
\item Quality `A' far side KDARs were awarded to \HII regions with statistically significant absorption in EIG features including at least the Near and Far 3kpc Arms.
\item Quality `B' far side KDARs were awarded to \HII regions with statistically significant absorption in any EIG feature located on the far side of the GC.
\item Quality `C' far side KDARs were awarded to \HII regions with large uncertainties ($>50\%$) in their calculated $D_{los}$ value (see \S \ref{errors}).
\end{itemize}
\item NEAR: If the \HI absorption spectrum displays no absorption associated with any EIG features, then it must be located at the `near' kinematic location.  Here we assume that all the EIG features are visible within the latitude range of the target \HII regions ($|b|<1.5^{o}$).  Note that the linear scale heights of the Near and Far 3kpc Arms is $\sim103$ pc FWHM \citep{DameThaddeus08}; assuming that the Far 3kpc Arm is at a uniform line of sight distance of 11.5kpc, this scale height corresponds to a latitude range $|b|\lesssim0.5^{o}$.  As a result, if a \HI absorption spectrum towards an \HII region located at $|b|>0.5^{o}$ displays no absorption associated with any EIG feature, it is awarded a Near KDAR, of Quality 'C'.
\begin{itemize}
\item KDARs of Quality `A' were given to all near side \HII regions, unless
\item the calculated $D_{los}$ value had large uncertainties ($>50\%$), then a Quality `C' KDAR was given.
\end{itemize}
\item No KDAR was attempted for \HII regions with multiple RRL velocity components, as multiple systemic velocities suggest several ionisation sources along the line of sight.  Note: multi-RRL velocity component \HII regions account for less than 10\% of the \HII regions within $|l|<10^{o}$, compared with 30\% for the Galactic plane in general (GBTHRDS).
\end{itemize}

\HII regions with calculated kinematic distances are listed in Table \ref{Table4}.

Four \HII regions (G350.177+00.017, G350.330+00.157, G353.557-00.014 and G003.949-00.100) were deemed to lie at the far kinematic location, beyond the EIG, following the rules above.  However, these four regions have $R_{Gal}<3$kpc, i.e. they are outside the bounds of the \citet{MCGD07} rotation model and are therefore not included in Table \ref{Table4}.

\begin{table}[h!]
\centering
\begin{tabular}{llcccl}
\hline
\hline
\HII Region& V$_{\text{RRL}}$&Ref.&Q&$R_{Gal}$&$D_{los}$\\
\hline
\multicolumn{6}{l}{KDAR: Far - within rotation model boundaries}\\
\hline
G$350.813-00.019$&	$+0.3$	&3&	B&	8.6&			16.9	$\pm$4.2\\
G$351.028+00.155$&	$+4.8$	&1&	A&	9.7&			18.0	$\pm$5.9\\
G$351.358+00.666$&	$-3.6$	&3&	B&	8.2&			16.5	$\pm$3.6\\
G$352.313-00.440$&	$-13.4$	&1&	A&	5.9&			14.3	$\pm$2.6\\
G$354.486+00.085$&	$+15.8$	&3&	C&	8.8&			23.3	$\pm$*\\	
G$354.610+00.484$&	$-23.4$	&1&	A&	3.9&			12.3	$\pm$2.2\\
G$354.934+00.327$&	$+14$	&4&	C&	14.6&		23.1	$\pm$*\\
G$355.242+00.096$&	$+10.3$	&3&	C&	13.3&		21.7	$\pm$*\\
G$355.344+00.145$&	$+16.4$	&1&	C&	16.8&		24.8	$\pm$*\\
G$000.640+00.623$&	$+3.7$	&2&	C&	3.3&			11.8	$\pm$*\\
G$000.838+00.189$&	$+5.6$	&1&	C&	3.0&			11.5	$\pm$*\\
G$003.270-00.101$&	$+4.9$	&3&	C&	6.3&			14.7	$\pm$*\\
G$003.342-00.079$&	$+8.3$	&3&	A&	5.2&			13.6	$\pm$6.7\\
G$004.412+00.118$&	$+4.1$	&3&	C&	7.1&			15.6	$\pm$8.9\\
G$004.527-00.136$&	$+10.2$	&1&	A&	5.4&			13.8	$\pm$3.9\\
G$005.479-00.241$&	$+21.4$	&3&	A&	4.1&			12.5	$\pm$2.3\\
G$005.524+00.033$&	$+23.3$	&1&	A&	4.0&			12.3	$\pm$2.2\\
G$006.083-00.117$&	$+8.8$	&3&	A&	6.3&			14.7	$\pm$3.5\\
G$006.553-00.095$&	$+15$	&3&	A&	5.3&			13.7	$\pm$2.5\\
G$006.565-00.297$&	$+20.9$	&3&	B&	4.6&			12.9	$\pm$2.2\\
G$007.041+00.176$&	$+8.9$	&1&	A&	6.6&			14.9	$\pm$3.1\\
G$007.420+00.366$&	$-12.3$	&1&	C&	12.2&		20.6	$\pm$*\\
G$008.006-00.156$&	$+42.6$	&3&	A&	3.4&			11.6	$\pm$1.9\\
G$008.432-00.276$&	$+34.8$	&1&	A&	4.0&			12.2	$\pm$2.0\\
\hline
\multicolumn{6}{l}{KDAR: Near}\\
\hline
G$351.192+00.708$&	$-3.4$	&3&	C&	8.2&		0.3$\pm$*\\
G$353.186+00.887$&	$-4.7$	&3&	C&	7.6&		0.9$\pm$*\\
G$359.277-00.264$&	$-2.4$	&5&	C&	4.5&		4.0$\pm$*\\
G$005.889-00.427$&	$+10.1$	&3&	C&	6.0&		2.5$\pm$*\\
G$008.137+00.228$&	$+20.6$	&3&	C&	5.1&		3.5$\pm$2.2\\
G$008.666-00.351$&	$+49.1$	&3&	A&	4.7&		3.9$\pm$1.4\\
G$008.830-00.715$&	$+26.6$	&1&	C&	4.7&		3.9$\pm$2.0\\

\hline 
\hline
\end{tabular}
\caption{\HII regions with calculated kinematic distances (in kpc).  References for the RRL velocities are as follows: 1) GBTHRDS (2011), 2) \citet{LPH96}, 3) \citet{Lockman89}, 4) \citet{CH87}, 5) \citet{WAM82}.  Errors in kinematic distances marked with asterisks (*) denote errors which are $>100\%$; note that while these errors are large, the significance of the KDAR remains.} 
\label{Table4}
\end{table}

If an \HII region is awarded a far side KDAR, based on the above requirements, a kinematic distance can be calculated using a Galactic rotation curve model (which assumes circular rotation).  The IAU Galactic Constants have been applied in these calculations: $R_{0}=8.5 \text{ kpc and } \Theta_{0}=220$\kms.

If an \HII region must be located at least as far as the subcentral point, then its location inside, or beyond, the Solar circle is given by its systemic velocity.  In the Inner Galaxy, velocities are positive in the first quadrant and negative in the fourth.  The signs are reversed in the Outer Galaxy, such that first quadrant sources located beyond the Solar circle will have negative velocities, and fourth quadrant sources in the Outer Galaxy will have positive velocities.

Throughout this work, the rotation curve of \citet{MCGD07} is used for regions within the Solar Circle.  In the outer Galaxy, $D_{los}$ was calculated using a flat rotation model $\Theta_{R_{Gal}}=\Theta_{0}$.

\subsubsection{Kinematic Distance Uncertainties\label{errors}}

\begin{figure}[h!]
\centering
\includegraphics[width=\columnwidth]{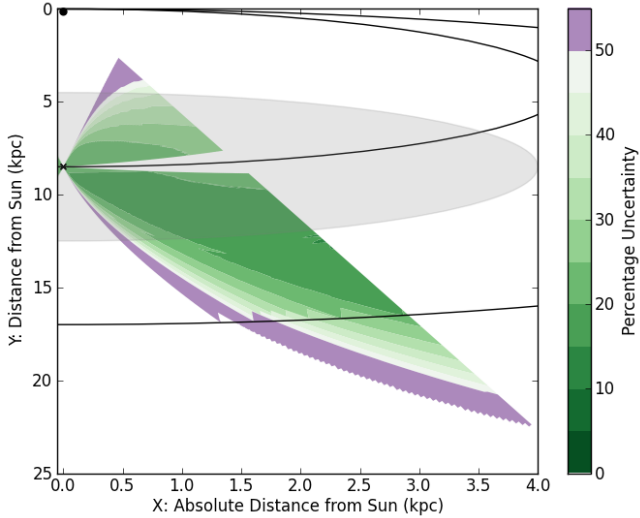}
\caption{Total percentage uncertainty in the line of sight distance $D_{los}$ caused by the choice of rotation curve, non-circular streaming motions of 15\kms and by changing the Solar circular rotation speed to $\Theta_{o}=255$\kms.  Blank areas are indicative of $|l|>10^{o}$ (where no error analysis was carried out), or percentage uncertainties $>100\%$. The EIG is shaded grey (no kinematic distances were calculated for this region), and the Solar Circle and Locus of Tangent Points appear as the black circles.  The percentage uncertainties are mirrored for lines of sight in the fourth quadrant, here, only $l>0$ is shown for clarity. (A color version of this figure is available in the online journal.) \label{Fig6}}
\end{figure}

We follow the distance uncertainty analysis method of \citet{Anderson12}, investigating the effects of the choice of rotation curve, streaming motions and Solar rotation speed on kinematic distance calculations.  We compare all sources of uncertainty to the distances calculated from the rotation model of \citet{MCGD07}.

Firstly we compute, for a grid of ($l,v$) positions, the kinematic distance using the rotation curves of \citet{MCGD07, BB93} and \citet{Clemens85}.  The standard deviation of these distances for each ($l,v$) is then computed and divided by the \citet{MCGD07} distance to obtain the percentage uncertainty due to choice of rotation curve.  We assessed a grid of $|l|<10^{o}$ and $|v|<200$\kms with steps of 0.1 in each unit.

This procedure is then repeated, but instead of varying the Galactic rotation model, the percentage uncertainty due to streaming motions (of 15\kms) and an altered circular Solar rotation speed of 255\kms \citep{Reid09} are investigated.

The effect of these three sources of error are then added in quadrature for each ($l,v$) pair, before transformation onto a face-on plot (Figure \ref{Fig6}).

It should be noted that kinematic distance errors at small Galactic longitudes are intrinsically larger than for other sections of the Galaxy.  Equal steps in velocity map to unequal length steps in $D_{los}$, such that $dD_{los}/dv\propto \sin^{-1} l$.  This can be seen in Figure \ref{Fig6} where the percentage uncertainty is higher for smaller longitudes.

\paragraph*{Uncertainties due to Rotation Curve}
In addition to larger uncertainties at small longitudes due to the velocity gradient, errors are also larger in the Outer Galaxy due to the uncertainty in the outer Galaxy circular rotation models.  Flat, rising and falling rotation curves have been suggested for beyond the Solar circle \citet{BB93,Honma07,Hachisuka09}.  
Here we have used the rotation models of \citet{MCGD07, BB93} and \citet{Clemens85}.  Note that the \citet{MCGD07} model has an applicable Galactocentric range of $3\leq R_{Gal}\leq 8$ kpc.  As a result the model was extrapolated to the Solar circle, and a flat rotation curve was assumed for $R_{Gal}>R_{o}$.  The largest discrepancy between these three models occurs at $R_{Gal}\sim10$ kpc, but even at $R_{Gal}<Ro$ the models differ significantly.

\paragraph*{Uncertainties due to Non-Circular, Streaming Motions}
Large non-circular motions have precluded Galactic astronomers from fitting a rotation curve to the EIG. Smaller-scale non-circular motions are ubiquitous in the Galaxy and are the result of systematic velocity fields within a source, or ordered large-scale Galactic streaming motions \citet{Anderson12}.  \citet{BaniaLockman84} suggest an uncertainty, due to non-circular motions, of 5 to 10\kms; whereas \citet{Kolpak03} assign an estimate of cloud-cloud dispersion of 5 \kms in addition to contributions from Galactic scale streaming motions of 10 \kms . \citet{Dickey03} and \citet{Jones12} find \HI absorption components extending to 10-20 \kms beyond the systemic velocity of \HII regions.

In order to promote a conservative approach to kinematic distance uncertainties, the random uncertainty due to non-circular motions is set to 15 \kms.  The contribution of errors due to streaming motions in relation to the total uncertainty in kinematic distance is high, especially for small longitudes. 
The errors due to non-circular motions are the standard deviation of the three $(l,v)$ grids,$(l,v)$, $(l,v+15)$, $(l,v-15)$ divided by the $(l,v)$ distance, all computed with the \citet{MCGD07} rotation model.

\paragraph*{Uncertainties due to Solar Rotation Parameters}
The IAU values for $R_{o}=8.5$ kpc and $\Theta_{o}=220$\kms have been used throughout this work.  However, here we investigate the significance of an altered Solar rotation speed, as suggested by \citet{Reid09}.
Two $(l,v)$ grids were computed with the rotation model of \citet{MCGD07}, using $\Theta_{o}=220,250$\kms.  The standard deviation of these two grids, at each locus, was then divided by the standard (i.e. $\Theta_{o}=220$\kms) distance to compute the percentage uncertainty due to choice of Solar rotation parameters.  Note that the \citet{Reid09} value for $R_{o}=8.4\pm0.6$ kpc is consistent with the IAU value, and is therefore not investigated here.

\subsection{Galactic Distribution of \HII Regions}
\begin{figure*}[bh!]
\centering
\includegraphics[width=0.47\textwidth]{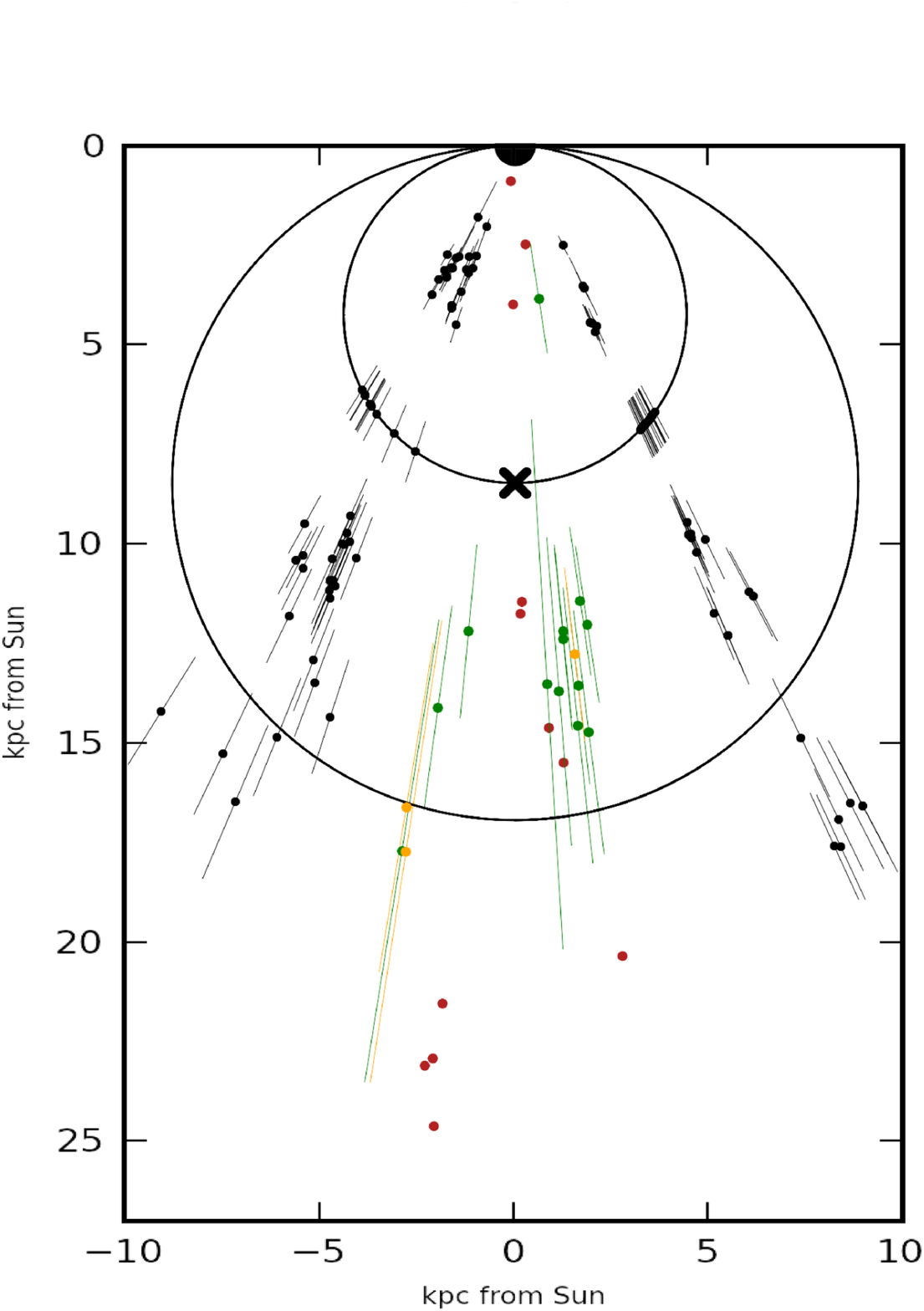}
\includegraphics[width=0.47\textwidth]{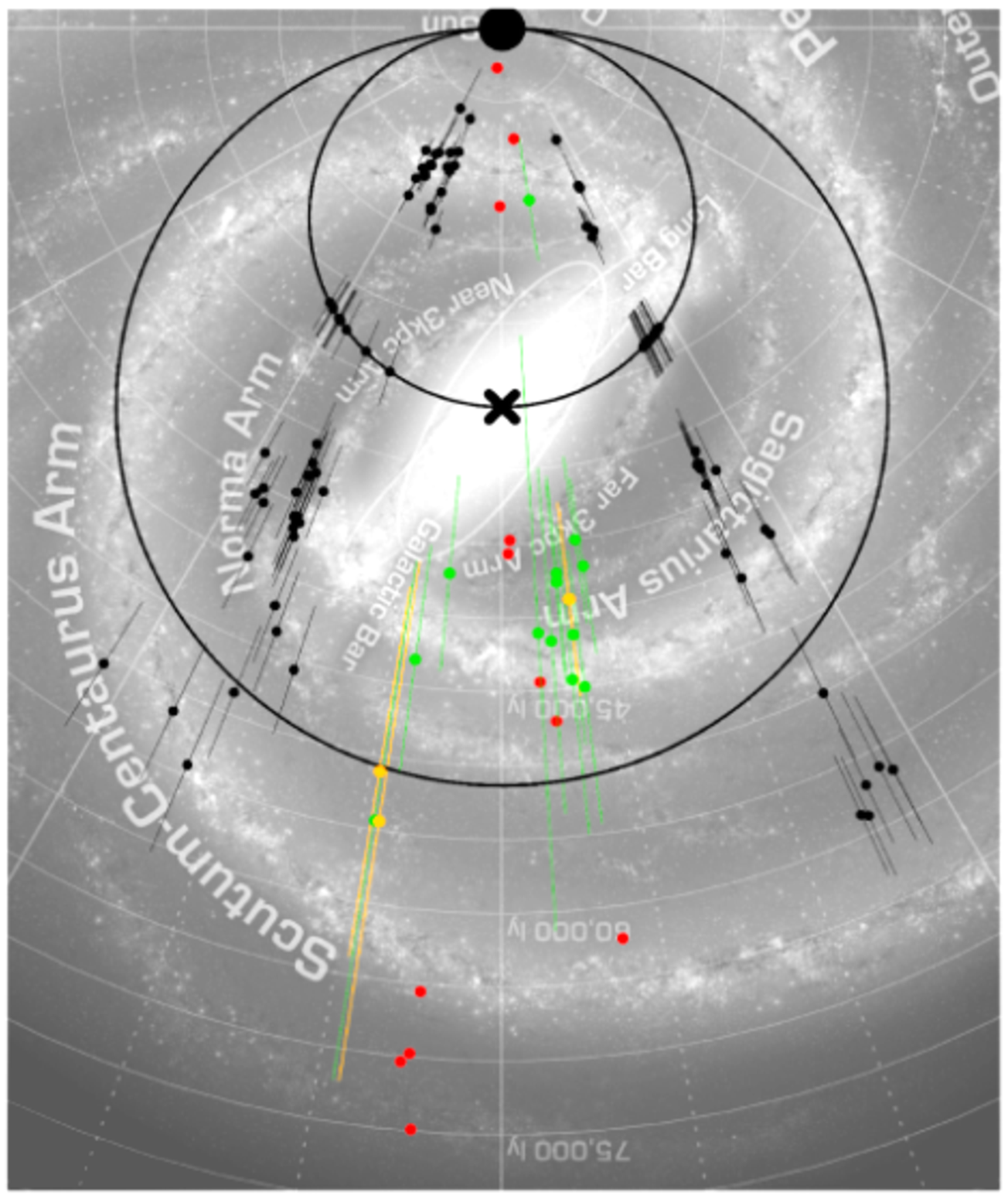}
\caption{Positions of the \HII region complexes for which a kinematic distance was calculated (Quality A, B and C shown as green, orange and red markers respectively).  Also shown are the Solar Circle and locus of tangent points (black circles) and kinematic distances for \HII regions from \citet{Jones12} (black markers).  Error bars are calculated according to the analysis of \S \ref{errors}; the large uncertainties are not shown for quality C distances.  Background image credit [right panel]: Hurt \& Benjamin in \citet{GLIMPSE}. (A color version of this figure is available in the online journal.) \label{Fig7}}
\end{figure*}

In order to examine the large-scale structure of the Galaxy, \HII regions with successfully calculated kinematic distances were transformed into a face-on map of the Milky Way (left panel of Figure \ref{Fig7}) and also superimposed onto an artist's conception of the Galaxy (right panel).  The background image used in the right panel of Figure \ref{Fig7}, was created using stellar, \HI and CO data \citep{GLIMPSE} and was reviewed in \citet{Urquhart12}.

In addition, the kinematic distances from \citet{Jones12} are also displayed.  Figure \ref{Fig7} demonstrates the need for \HII region discovery and KDAR studies for Galactic longitudes $340^{o}<l<350^{o}$ in order to further investigate the end of the bar and differentiate the Norma and Near 3kpc Arms, as well as the Sagittarius and Perseus Arms on the far side of the locus of subcentral points (smaller circle in Figure \ref{Fig7}).  At the end of the bar in the first quadrant, \HII region KDARs have been made by \citet{AndersonBania09} and \citet{Bania12} - further encouragement for a fourth quadrant study.

Recently, \citet{DameThaddeus11} identified an extension of the Scutum-Centaurus Arm at extreme distances from the Sun, in the first Galactic quadrant.  However, confirmation of this discovery requires tracing the Arm over its entire longitude range.  \citet{DameThaddeus11} comment that molecular gas which constitutes the section of Scutum-Centaurus Arm behind the Galactic center will be the most difficult to deconvolve.  In the longitude range of this paper, \HII regions with systemic velocities opposite in sign to circular-disk rotation must be located in the EIG or beyond the Solar circle.  Using \HI absorption features to resolve this ambiguity has allowed for seven \HII regions to be unequivocally placed in the outer Galaxy.  Several of these outer Galaxy regions (see Figure \ref{Fig7}) appear to trace the Scutum-Centaurus arm.

\section{Summary}
The EIG remains a difficult section of the Milky Way to study.  In terms of Galactic structure, kinematic studies in this region are hampered by a lack of rotation model for $R_{Gal}<4$kpc (and for the outer Galaxy).  In addition, there remains a lack of consensus regarding the number, locations and nomenclature of large-scale structures near the Galactic Centre (these are discussed in Section \ref{GC}).  Despite this, \HI absorption associated with EIG features was successfully used as a distance indicator, allowing for constraints on the line of sight distance for over 80\% of the sample of \HII regions investigated, or over 60\% of all known \HII regions with systemic velocities in $|l|<10^{o}$.

Over 67\% of the \HII regions demonstrate \HI absorption associated with the Near 3kpc Arm (see Table \ref{Table2}) and therefore must be located at line of sight distances of at least $\sim5$kpc.  A further 16 \HII regions show absorption associated with EIG features assumed to lie further along the line of sight than the Near 3kpc Arm, therefore, over 78\% of the sample \HII regions are located at $D_{los}\gtrsim 5$kpc.  This is in keeping with the work of \citet{Lang10} who find $\sim90\%$ of their sample of 40 EIG continuum sources must be located at least as far as the Near 3kpc Arm.

Of the 151 \HII regions investigated, 54 \HII regions display absorption from EIG features assumed to be on the far side of the GC (the +135\kms Expanding Arm, Bania's Clump 1 or Far 3kpc Arm).  Consequently, these \HII regions must be located at $D_{los}\gtrsim 8.5$kpc.

After successfully resolving the near/far kinematic distance ambiguity, line of sight distances were calculated for 31 \HII regions.  These distances suggest locations for the \HII regions in known Galactic structures including the Norma, Sagittarius and Perseus spiral Arms (see Figure \ref{Fig7}).  The 7 \HII regions beyond the Solar Circle are among the most distant Galactic \HII regions known to exist and could be crucial to tracing the Scutum-Centaurus Arm; where identification of star formation with molecular tracers is extremely difficult \citep{DameThaddeus11}.  Errors on these line of sight distances are often large - due to the uncertainty of non-circular streaming motions, and differences in Galactic rotation models - but the near/far KDAR remains both valid and significant.

Using a summary of EIG structures, and the known $lv$ distribution of CO, we construct a `crayon diagram' with which to investigate the distribution of \HI absorption in the EIG (Figure \ref{Fig2}, Section \ref{cayondiagramsection}).  In Section \ref{HIdist} we find cold \HI clouds, signified by \HI absorption, associated with the Near 3kpc Arm, Connecting Arm, Bania's Clump 1, Tilted Disk and Far 3kpc Arm.  There was minimal \HI absorption associated with either the Looping Ridge or the +135\kms Expanding Arm.  The large amount of \HI absorption associated with each of the 3kpc Arms presented an opportunity to fit a model to the $lv$ locus of each Arm (\S \ref{HIin3kpcArms}).  We find a linear fit (in $lv$ space) that is consistent with the findings of \citet{DameThaddeus08}, who used CO to trace the Arms.

The $lv$ distribution of the RRL velocities of the 151 \HII regions was investigated in Section \ref{HIIdist}.  Like the \HI absorption distribution, the systemic velocities of the \HII regions trace Galactic structures including spiral arms, features located near the Galactic center and possibly the end of the bar.  While most \HII regions posses RRL velocity components allowed by circular Galactic rotation (suggestive of a location outside the EIG), smaller numbers of \HII regions are found to be associated with the \HI Tilted Disk, Near 3kpc Arm, +135\kms Expanding Arm, Bania's Clump 1, Connecting Arm and Far 3kpc Arm.  Using the RRL velocity and \HI absorption spectrum of each \HII region, we were also able to constrain the $D_{los}$ for a further sample of \HII regions using only EIG features as a distance indicator.

\acknowledgments
This research has made use of NASA's Astrophysics Data System, the NASA/IPAC Extragalactic Database (NED) and the SIMBAD database. 
\clearpage

%% Appendix material should be preceded with a single \appendix command.
%% There should be a \section command for each appendix. Mark appendix
%% subsections with the same markup you use in the main body of the paper.

%% Each Appendix (indicated with \section) will be lettered A, B, C, etc.
%% The equation counter will reset when it encounters the \appendix
%% command and will number appendix equations (A1), (A2), etc.

\appendix
\section{Discussion of Individual \HII Regions}
\label{individual}
\subsection*{G350.004+00.438}
The \HI absorption spectrum does not give a clear indication of any absorption associated with any EIG feature.  At this longitude, the velocity range of the Far 3kpc Arm is not clearly distinct from velocities expected by normal circular rotation. 
\subsection*{\objectname[GAL 350.13+00.09]{G350.129+00.088}}
The \HI absorption spectra clearly demonstrates absorption either side of the velocities expected by an association with the Near 3kpc Arm.  \citet{Quireza06} place the \HII region at a line of sight distance of 6.2 kpc (i.e. on the near side of the GC). 
\subsection*{G350.177+00.017}
Evidence of absorption in the Far 3kpc Arms suggests a far KDAR.  As with G350.330+00.157 (below), the calculated $R_{Gal}$ and $D_{los}$ for the region are outside the bounds of the \citet{MCGD07} rotation model.
\subsection*{G350.330+00.157}
While the \HI absorption spectrum suffers from emission fluctuations around the RRL velocity ($\sim-60$\kms), there is evidence of absorption associated with the Near 3kpc Arm.  Assuming a far side KDAR, the calculated $R_{Gal}$ and $D_{los}$ are outside the bounds of the \citet{MCGD07} rotation model.
\subsection*{G350.335+00.107}
Evidence of \HI absorption is seen either side of the velocities associated with the Near 3kpc Arm (see G350.129+00.088 above), and is therefore located at least as far as the Near 3kpc Arm along the line of sight.
\subsection*{\objectname[GAL 350.52+00.96]{G350.524+00.960}}
G350.524+00.960 does not demonstrate any \HI absorption outside the velocities expected by normal circular rotation.  If the near kinematic distance is therefore assumed, the \HII region has a calculated $D_{los}\approx1.9$kpc.
\subsection*{G350.769-00.075}
The \HI absorption spectrum of G350.769-00.075 does not give conclusive evidence for either a near, nor far, KDAR.
\subsection*{\objectname[GAL 350.81-00.02]{G350.813-00.019}}
As the \HI absorption spectrum of G350.813-00.019 demonstrates absorption in velocities associated with the Far 3kpc Arm, the \HII region must be on the far side of the GC.  The positive (small) RRL velocity then locates the \HII region at a line of sight distance beyond (but close to) the Solar Circle.
\subsection*{\objectname[GAL 351.00-00.58]{G350.996-00.557}}
Strong absorption is seen in the Far 3kpc Arm, but not in the Near 3kpc Arm.  The RRL velocity suggests a location within the Far 3kpc Arm.
\subsection*{G351.028+00.155}
The \HI absorption spectrum of G351.028+00.155 demonstrates significant absorption at velocities corresponding to both the Near and Far 3kpc Arms, and therefore must be located at least as far as the Far 3kpc Arm along the line of sight.  The positive RRL velocity then requires that G351.028+00.155 is located in the outer Galaxy.
\subsection*{G351.047-00.322}
The \HI absorption spectrum of G351.047-00.322 does not give conclusive evidence for either a near, nor far, KDAR.
\subsection*{\objectname[L89b 351.192+00.708]{G351.192+00.708}}
\HI absorption is evident in circular rotation allowed velocities only. If the near kinematic distance is therefore assumed, the \HII region has a calculated $D_{los}\approx0.3$kpc.  \citet{Moises11} assumes the near kinematic distance, however \citet{Quireza06} place the \HII region at a line of sight distance of 17.1 kpc.
\subsection*{\objectname[L89b 351.201+00.483]{G351.201+00.483}}
The \HI absorption spectrum of G351.201+00.483 does not give conclusive evidence for either a near, nor far, KDAR. \citet{Quireza06} place the object at 1.4 kpc, at the near kinematic location.
\subsection*{\objectname[GAL 351.36+00.67]{G351.358+00.666}}
\citet{Quireza06} assumes a near KDAR for G351.358+00.666, but \HI absorption associated with velocities expected of the Far 3kpc Arm suggest that the \HII region is located at the far kinematic location.  
\subsection*{G351.359+01.014}
The \HI absorption spectrum of G351.359+0.1014 does not give conclusive evidence for either a near, nor far, KDAR.
\subsection*{G351.467-00.462}
\citet{Quireza06} give G351.467-00.462 a near side KDAR, but the \HI absorption spectrum from this paper does not give conclusive evidence for a KDAR.
\subsection*{\objectname[GAL 351.60-00.35]{G351.601-00.348}}
The RRL velocity for this \HII region (-91.8 \kms, \citet{Lockman89}) is associated with the Near 3kpc Arm.  \citet{maserKDA} also position a nearby 6.7 GHz maser ($l,b=351.581,-0.353$)in the Near 3kpc arm.
\subsection*{\objectname[GAL 351.66+00.52]{G351.662+00.518}}
G351.662+00.518 has a near zero RRL velocity (-2.9\kms, \citet{Lockman89}) which is associated with locations inside the EIG region, near the Solar Cirle, or at a very small line of sight distance from the Sun.  Absorption at velocities associated with the Near 3kpc Arm imply a $D_{los}>5$kpc.  As there is no \HI absorption associated with other EIG features (only the Far 3kpc Arm is expected at this longitude), a location within $R_{Gal}\lesssim3$kpc is assumed.
\subsection*{G351.691+00.669}
No \HI absorption falls outside the circular rotation envelope of allowed velocities, suggesting a near KDAR.  However, the positive RRL velocity suggests a location in either the EIG or outer Galaxy.  
\subsection*{G352.233-00.151}
This \HII region has an RRL velocity associated with the Near 3kpc Arm (-88.6\kms, GBTHRDS).  Strong absorption in the allowed circular rotation velocities and at velocities associated with the Near 3kpc Arm, reaffirm the location in the Arm.
\subsection*{G352.313-00.440}
Evidence of \HI absorption in both the Near and Far 3kpc Arms suggests a far side KDAR for G351.313-00.440.
\subsection*{\objectname[GAL 352.40-00.06]{G352.398-00.057}}
Absorption is seen at the expected velocities of the Near 3kpc Arm, which is also where the RRL velocity for this \HII region lies (-87\kms, \citet{Lockman89}).  Absorption up to 25 \kms beyond the RRL velocity of an \HII region is not uncommon \citep{Dickey03, Jones12}, therefore it is assumed that the \HII region is located in the Near 3kpc Arm.
\subsection*{G352.521-00.144}
Two RRL velocities have been recorded for G352.521-00.144 (-57.3 and -38 \kms, GBTHRDS), suggestive of multiple emission sources along the line of sight.
\subsection{G352.610+00.177}
The \HI absorption spectrum for G352.610+00.177 suffers from emission fluctuations.  As a result the poor quality spectrum does not give conclusive evidence for a KDAR.
\subsection*{\objectname[GAL 352.61-00.17]{G352.611-00.172}}
G352.611-00.172 displays strong absorption at $\sim100$\kms, approximately 20\kms beyond the known RRL velocity of the \HII region (-81.9\kms, \citet{Lockman89}).  As with G352.398-00.057 (above), G352.611-00.172 is assumed to lie in the Near 3kpc Arm.  This location, in the Near 3kpc Arm, is approximately the same as the line of sight distance given by \citet{Quireza06} (6.7 kpc).
\subsection*{\objectname[GAL 352.87-00.20]{G352.866-00.199}}
Evidence of absorption at velocities corresponding to the Near 3kpc Arm suggest a $D_{los}\geq5$kpc.  \citet{maserKDA} position a nearby 6.7 GHz methanol maser  ($l,b=352.855,-0.201$) at the far kinematic location ($D_{los}\approx11$kpc). 
\subsection*{\objectname[L89b 353.186+00.887]{G353.186+00.887}}
\HI absorption is evident in circular rotation allowed velocities only, G353.186+00.887.  If the near kinematic location is then assumed, the \HII region has a calculated $D_{los}\approx0.9$kpc.  \citet{Quireza06} provide a near side KDAR for this \HII region.
\subsection*{G353.218-00.249}
Also the source of a variable maser \citep{Caswell10}, G353.218-00.249 has a small RRL velocity (-8.3 \kms, GBTHRDS) and absorption present at Near 3kpc Arm, but not Far 3kpc Arm, velocities.  These are evidence for a location near the EIG, and as such $D_{los}\geq5$kpc, $R_{gal}<3$kpc is assumed.
\subsection*{\objectname[GAL 353.38-00.11]{G353.381-00.114}}
The \HI absorption spectrum of G353.381-00.114 displays strong \HI absorption associated with the Near 3kpc Arm, suggesting that the \HII region must lie behind the feature.
\subsection*{\objectname[GAL 353.398-00.3]{G353.398-00.391}}
\HI absorption is evident in circular rotation allowed velocities only, if, therefore, a near side KDAR is assumed, the calculated $D_{los}\approx5.2$kpc.
\subsection*{\objectname[GAL 353.56-00.01]{G353.557-00.014}}
\HI absorption is present at velocities corresponding to both 3kpc Arms, suggestive of a far side KDAR.  However, like the \HII regions G350.330+00.157 and G350.177+00.017, the calculated $R_{Gal} \text{ and } D_{los}$ fall outside the boundaries of the \citet{MCGD07} rotation model.
\subsection*{\object[GAL 354.20-00.05]{G354.200-00.050}}
Strong absorption is centered at velocities to the negative side of those expected for the Near 3kpc Arm (see G352.611-00.172 and G352.398-0.057 above).\citet{maserKDA} were unable to determine a KDAR for a nearby 6.7 GHz methanol maser ($l,b=354.206,-0.038$).  Due to the \HI absorption associated with the Near 3kpc Arm $D_{los}\geq5$kpc is assumed.
\subsection*{G354.418+0.036}
The \HI absorption spectrum of G354.418+0.036 does not give conclusive evidence for either a near, nor far, KDAR.
\subsection*{\objectname[GAL 354.49+00.09]{G354.486+00.085}}
\citet{Caswell10} places a nearby 6.7 GHz methanol maser ($l,b=354.496,0.083$) in the Far 3kpc Arm.  The RRL velocity of the \HII region (15.8\kms, \citet{Lockman89}) is slightly smaller than that expected for the Far 3kpc Arm, but the absorption indicates the \HII region must be located at least as far along the line of sight as the feature.  Due to the positive RRL velocity, we assume that G354.486+00.085 is located beyond the Solar Circle (see Table \ref{Table4}).
\subsection*{G354.588+00.007}
A line of sight along the longitude of 354.588$^{o}$ intersects the Near and Far 3kpc Arms as well as the assumed position of Bania's Clump 1.  The \HI absorption spectrum of G354.588+00.007 does not give conclusive evidence for either a near, nor far, KDAR; but absorption associated with the Near 3kpc Arm suggests $D_{los}\geq5$kpc.
\subsection*{G354.610+00.484}
Significant \HI absorption is present before and after the velocities expected of the Near 3kpc Arm, as well as at Far 3kpc Arm velocities.  A known strong 6.7 GHz methanol maser is also in the region \citep{Caswell10}, with a velocity equivalent to the RRL velocity (maser velocity: -23\kms, RRL velocity: -23.4\kms (GBTHRDS)).  \citet{maserKDA} suggest a poor quality near side KDAR for the associated maser, but a far kinematic distance is assumed here.
\subsection*{\objectname[GAL 354.66+00.47]{G354.664+00.470}}
\HI absorption is evident in circular rotation allowed velocities only, assuming a near side KDAR the calculated $D_{los}\approx4.5$kpc.
\subsection*{\objectname[GAL 354.67+00.25]{G354.665+00.247}}
No absorption is seen at the RRL velocity of the \HII region (97.8\kms, \citet{Lockman89}), nor at velocities corresponding to the Near 3kpc Arm.  However, significant absorption is seen at $\sim70$\kms, possibly associated with Bania's Clump 1.  No KDAR is given here, however the high RRL velocity is suggestive of a location in the EIG \citep{Caswell10}.
\subsection*{G354.717+00.293}
As with G354.665+00.247, the high RRL velocity of G354.717+00.293 suggests a location in the EIG.  The \HI absorption spectrum suffers from emission fluctuations at the RRL velocity (95.3\kms, GBTHRDS) and no absorption is present at Near 3kpc Arm velocities.  At least two masers with high velocities ($\sim100$\kms) are known to exist in the area \citep{Caswell10}.
\subsection*{\objectname[GAL 354.93+00.33]{G354.934+00.327}}
G354.934+00.327 shares a similar absorption profile to that of G354.717+00.293 and G354.665+00.247, however it does not share a highly non-circular RRL velocity (14\kms, \citet{CH87}).  Absorption velocities corresponding to all expected EIG features requires the \HII region to be located at least as far along the line of sight as the Far 3kpc Arm.  Due to the positive RRL velocity, the \HII region must then be located in the outer Galaxy, beyond the Solar circle along the line of sight.
\subsection*{G354.979-00.528}
The \HI absorption spectrum of G354.979-00.528 does not give conclusive evidence for either a near, nor far, KDAR.
\subsection*{\objectname[L89b 355.242+00.096]{G355.242+00.096}}
\HI absorption is present at velocities corresponding to the Near 3kpc Arm on the near side of the GC, and there is evidence of absorption on the far side of the GC due to the Far 3kpc Arm and +135\kms Expanding Arm.  A far side KDAR is given, but due to the positive RRL velocity, the \HII region must be located beyond the Solar Circle.
\subsection*{G355.344+00.145}
Absorption at the Near 3kpc Arm, +135\kms Expanding Arm and Far 3kpc Arm infer that the \HII region is located beyond the EIG along the line of sight.  The positive RRL velocity then places the \HII region beyond the Solar circle. There are several masers in the region which are assumed to lie within 3 kpc of the GC \citep[see $(l,b)=(355.343,+0.148),(355.344,+0.147) \text{ and } (355.346,+0.149)$ in][]{maserKDA}.
\subsection*{G355.532-00.100}
This region has four known RRL velocities (3.8, -22.5, -80.6 and -41.1 \kms, GBTHRDS), a strong indication that there are several emission sources along the line of sight.  Note that the RRL velocity $-80.6$\kms is associated with velocities expected of the Near 3kpc Arm. No KDAR is given.
\subsection*{G355.581+00.288}
Three RRL velocities are known towards the \HII region (+108.7, -76.1 and +11.7 \kms, GBTHRDS).  As with G355.532-00.100 (above), this is an indication of several sources along the line of sight.  No KDAR is given, however the RRL velocity $-76.1$\kms is associated with the Near 3kpc Arm and the RRL velocity component $+108.7$\kms is associated with the +135\kms Expanding Arm.
\subsection*{G355.611+00.382}
The near zero RRL velocity (-2.6 \kms, GBTHRDS) is indicative of a EIG location, or a location near the Solar circle (either very close or at a great distance from the Sun).  Absorption in velocities associated with the Near 3kpc Arm and +135\kms Expanding Arm, but not at velocities corresponding to far side EIG features prompts $D_{los}\geq8.5$kpc, $R_{gal}<3$ kpc to be given as a distance limit for the \HII region.  In contrast, \citet{maserKDA} presents a far side KDAR for a 6.7 GHz methanol maser at $l,b=355.666,+0.398$ which has a systemic velocity of $\sim-2$\kms.
\subsection*{G355.696+0.350}
Two RRL velocities (3 and -79.1 \kms, GBTHRDS) suggest multiple emission sources along the line of sight, at least one of which is associated with the Near 3kpc Arm (-79.1\kms RRL association). At this longitude, the velocities of the Near 3kpc Arm and the Looping Ridge (on the far side of the GC) overlap.  No KDAR is given.
\subsection*{G355.700-00.100}
G355.700-00.100 has an absorption profile and RRL velocity (-76.1 \kms, \citet{Lockman89}) suggestive of a location within the Near 3kpc Arm or Looping Ridge (as the expected velocities of these features overlap at this longitude).  No KDAR is given.
\subsection*{G355.734+0.138}
There are multiple RRL velocities associated with G355.734+0.138 (10.7 and -77.4 \kms, GBTHRDS).  No KDAR is given, but the RRL velocity component at $-77.4$\kms is associated with the velocities expected of the Near 3kpc Arm or Looping Ridge.
\subsection*{G355.801-00.253}
The velocity ranges of the Near 3kpc Arm and Looping Ridge continue to overlap at this longitude.  Two RRL velocities are known (-31.5, 3.1 \kms, GBTHRDS), suggestive of multiple sources along the line of sight.  No KDAR is given.
\subsection*{G356.230+00.066}
At this longitude the expected velocities of the Near 3kpc Arm and Looping Ridge are distinct (see above).  However, the \HI absorption spectrum of G356.230+00.066 does not give conclusive evidence for either a near, nor far, KDAR.
\subsection*{G356.235+00.642}
Absorption is seen at velocities corresponding to the Near 3kpc Arm and Looping Ridge.  It is assumed that the \HII region is located in the +135\kms Expanding Arm (due to the RRL velocity (116.3\kms, \citet{Lockman89}).  This is supported by absorption at velocities corresponding to the Looping Ridge (on the far side of the GC, but closer to the GC than the +135\kms Expanding Arm).
\subsection*{G356.307-00.210}
A near zero RRL velocity (-4\kms, \citet{Lockman89}) and absorption concurrent with Near 3kpc Arm velocities suggests $R_{gal}<3$ kpc for this \HII region.
\subsection*{G356.470-0.001}
The \HI absorption spectrum of G356.470-0.001 does not give conclusive evidence for either a near, nor far, KDAR.
\subsection*{G356.560-00.086}
The \HI absorption spectrum of G356.560-00.086 does not give conclusive evidence for either a near, nor far, KDAR.
\subsection*{\objectname[GAL 356.65+00.13]{G356.650+00.129}}
\HI absorption is present in velocities corresponding to the Near 3kpc Arm.  As such $D_{los}>5$kpc is assumed.
\subsection*{G357.484-00.036}
The \HI absorption spectrum suffers from emission fluctuations in the velocity ranges associated with the Near 3kpc Arm and Looping Ridge.  As such the poor quality spectrum does not allow a KDAR to be given for this \HII region.
\subsection*{G357.970-00.169}
The \HI absorption spectrum of G357.970-00.169 displays absorption at velocities associated with the Near 3kpc Arm.  As a result, $D_{los}>5kpc$ is assumed.  The small RRL velocity, and lack of absorption corresponding to other EIG features suggests a further constraint, $R_{Gal}<3$kpc.
\subsection*{\objectname[L89b 357.988-00.159]{G357.998-00.159}}
The \HI absorption spectrum of G357.998-00.159 displays absorption at velocities associated with the Near 3kpc Arm.  As a result, $D_{los}>5kpc$ is assumed.  The small RRL velocity, and lack of absorption corresponding to other EIG features suggests a further constraint, $R_{Gal}<3$kpc.
\subsection*{\objectname[LPH96 358.319-0.414]{G358.319-00.414}}
The \HI absorption spectrum of G358.319-0.414 does not give conclusive evidence for either a near, nor far, KDAR.
\subsection*{G358.379-00.840}
The \HI absorption spectrum of G358.379-00.840 does not give conclusive evidence for either a near, nor far, KDAR.
\subsection*{G358.530+00.056}
This \HII region has an RRL associated with the Looping Ridge or Tilted Disk(-208.5\kms, GBTHRDS), however the spectrum is of poor quality and no absorption is seen at velocities pertaining to any EIG feature.
\subsection*{G358.552-00.025}
This \HII region has an RRL associated with the Looping Ridge or Tilted Disk(-208.5\kms, GBTHRDS), however the spectrum is of poor quality and no absorption is seen at velocities pertaining to EIG features in front of the GC along the line of sight.
\subsection*{G358.616-00.076}
The \HII region has an RRL association with the Tilted Disk or Looping Ridge.  The \HI absorption spectrum confirms absorption at velocities corresponding to the Near 3kpc Arm only; further supporting a location in the EIG.  Absorption is also seen at velocities either side of the expected velocity range of the Tilted Disk.
\subsection*{\objectname[GAL 358.62-00.07]{G358.623-00.066}}
Like G358.616-00.076 (above), G358.623-00.066 demonstrates significant absorption associated with the Near 3kpc Arm and Tilted Disk.  The RRL association with the Looping Ridge/Tilted Disk suggests a location in the EIG.  Note the bifurcation in the Near 3kpc Arm absorption profile, see \S \ref{cayondiagramsection}.
\subsection*{G358.633+00.062}
\HI absorption is seen at velocities corresponding to the Near 3kpc Arm and the \HI Tilted Disk.  The positive RRL velocity suggests either a EIG or near Solar circle location: absorption corresponding to near-side EIG features discounts the near-kinematic distance; and if the \HII region was located near the Solar cirlce on the far side, there should be evidence of absorption associated with the Far 3kpc Arm.  As a result it is assumed that the \HII region is located within the EIG, i.e. $R_{Gal}<3$kpc, $D_{los}>8.5$kpc.  Note the bifurcation in the Near 3kpc Arm absorption profile (see G358.623-00.066 above).
\subsection*{G358.652-00.078, G358.680-00.087, G358.694-00.075, G358.720+00.011, \objectname[LPH96 358.797+0.058]{G358.797+00.058}, G358.827+00.085 and G359.159-00.038}
The \HI absorption profiles of these \HII regions are all similar.  And all have highly non-circular RRL velocities which correspond to the Tilted Disk - G359.159-00.038 has two known RRL velocities (-182.5 and -215.6 \kms).  The \HI absorption spectra suffer from emission fluctuations and are generally of poor quality.
\subsection*{\objectname[CKW87 174143.4-294018]{G359.277-00.264}}
G359.277-00.264 demonstrates no absorption at velocities corresponding to EIG features, a near side KDAR is given.
\subsection*{G359.432-00.086}
G359.432-00.086 has a known RRL velocity associated with the Near 3kpc Arm.  The \HI absorption spectrum towards the region demonstrates absorption associated with the Near 3kpc Arm (and also at $\sim-120$\kms).
\subsection*{G359.467-00.172}
At this longitude the expected velocity ranges of the Near 3kpc Arm and Tilted Disk overlap.  The \HI absorption spectrum demonstrates absorption at velocities corresponding to the Near 3kpc Arm, but suffers from emission fluctuations at the overlap.  The \HII region has an RRL velocity consistent with either the Tilted Disk or Near 3kpc Arm.
\subsection*{\objectname[KC97c G000.3-00.5]{G000.284-00.478}}
Absorption is present at velocities corresponding to the Near 3kpc Arm, but not at the expected velocities of other EIG features.  It is assumed that the \HII region is located in the EIG, beyond the Near 3kpc Arm; $R_{Gal}<3$kpc, $D_{los}>8.5$kpc.
\subsection*{\objectname[KC97c G000.4-00.8]{G000.361-00.780}}
G000.361-00.780 demonstrates \HI absorption at velocities associated with the Near 3kpc Arm, but no absorption at other EIG $lv$ features.  It is therefore assumed, as with G000.284-00.478 above that the \HII region is located within $R_{Gal}<3$kpc, $D_{los}>8.5$kpc. 
\subsection*{G000.382+00.107}
With two RRL velocities (25.7 and 41.4 \kms, GBTHRDS), the absorption spectrum is likely to have contributions from at least emission two sources along the line of sight.  Absorption at the Near 3kpc Arm and +135\kms Expanding Arm velocities suggests that at least one of the emission sources is located on the far side of the GC.  No KDAR is given.  Note also that at this longitude, the expected velocity ranges of the \HI Tilted Disk and Far 3kpc Arm are nearly indistinguishable.
\subsection*{\objectname[NAME SGR B1]{000.510-00.051}}
\HI absorption is present at velocities corresponding to the Near 3kpc Arm, but not at the velocities of other EIG features.  The RRL velocity (45\kms, \citet{Downes80}) suggests an association with the Far 3kpc Arm.
\subsection*{G000.572-00.628}
The \HI absorption spectrum of G000.572-00.628 does not give conclusive evidence for either a near, nor far, KDAR.
\subsection*{\objectname[LPH96 000.640+0.623]{G000.640+00.623}}
A far-side KDAR is assumed for G000.640+00.623 due to absorption at velocities corresponding to both 3kpc Arms (and the Tilted Disk).
\subsection*{G000.729-00.103}
G000.729-00.123 has two recorded RRL velocities (105.3 and 83.2 \kms, GBTHRDS), both forbidden by circular Galactic rotation.  The region was studied by \citet{Downes80} who found an RRL velocity of 102 \kms.  \citet{CH87} discussed the \HII region as being clearly located near the EIG, but not delineating the outer boundary of the Galactic bar.  The GBTHRDS find that of their nine \HII regions associated (in $lv$ space) with the Nuclear Disk, G000.729-0.103 is the only source that could be located on the red-shifted side.  The \HI spectrum demonstrates statistically significant absorption at velocities corresponding to both 3kpc Arms, but not for the Nuclear Disk nor +135\kms Expanding Arm.  No \HI absorption is present at either of the RRL velocities.  No KDAR is given.
\subsection*{\objectname[KC97c G000.8+00.2]{G000.838+00.189}}
The \HI absorption spectrum, which is of poor quality due to emission fluctuations, demonstrates absorption consistent with the velocities expected of each of the EIG features (Near 3kpc Arm, \HI Tilted Disk, +135\kms Expanding Arm and Far 3kpc Arm).  A far-side KDAR is therefore awarded to the \HII region.
\subsection*{\objectname[NAME SGR D HII]{G001.125-00.105}}
\citet{WAM82} remarked that the non-circular RRL velocity (-19.7 \kms) and H$_{2}$CO at 84 and 123 \kms was typical of a EIG source; \citet{Quireza06} also give $D_{los}=8.5$kpc.  THe \HII region must be located within the EIG, as absorption at EIG features negates the near-side kinematic location and if the \HII region must have a non-realistic $R_{Gal}>45$kpc.
\subsection*{\objectname[KC97c G001.1-00.1]{G001.149-00.062}}
G001.149-00.062 displays absorption at velocities corresponding to both the Near and Far 3kpc Arms.  Assuming a distance of at least the Far 3kpc Arm, G001.149-00.062 must lie in the outer Galaxy, beyond the Solar Circle (due to the negative systemic velocity).  However, using a flat rotation model for the outer Galaxy, the calculated $D_{los}$ is unrealistic ($\sim50$kpc).  Therefore the \HII region must lie in the EIG region, but behind the Far 3kpc Arm.
\subsection*{G001.324+00.104}
No \HI absorption is seen at velocities corresponding to EIG features, suggesting a near KDAR. However, the negative RRL velocity (-12.7 \kms, GBTHRDS) suggests a location in either the EIG or in the outer Galaxy - locations that each would imply absorption by the Near 3kpc Arm, which is not seen.  No KDAR is given.
\subsection*{\objectname[GAL 001.32+00.09]{G001.330+00.088}}
G001.330+00.088 has a similar \HI absorption profile as G001.324+00.104.  A EIG location is assumed.
\subsection*{\objectname[L92 Sgr D 9]{G001.488-0.199}}
\citet{Caswell10} assigns a 6.7 GHz methanol maser at the same velocity to $R_{gal}<3$ kpc due to the negative systemic velocity.  Absorption at velocities corresponding to the Near 3kpc Arm supports the $R_{gal}<3$ kpc placement.
\subsection*{G002.009-0.680}
$D_{los}>5$ kpc is assumed due to absorption at Near 3kpc Arm velocities.
\subsection*{G002.404+0.068}
The \HI absorption spectrum of G002.404+0.068 does not give conclusive evidence for either a near, nor far, KDAR.
\subsection*{G002.418-0.611}
The \HI absorption spectrum of G002.418-0.611 does not give conclusive evidence for either a near, nor far, KDAR.
\subsection*{\objectname[LPH96 002.510-0.028]{G002.510-00.028}}
$D_{los}>5$kpc is assumed due to absorption at velocities corresponding to the Near 3kpc Arm.
\subsection*{\objectname[IRAS 17480-2636]{G002.611+00.135}}
For a 6.7 GHz methanol maser at the same coordinates, \citet{Caswell10} discuss that the large positive systemic velocity is most readily attributed to a location within the Galactic bar.  Absorption is seen at Near 3kpc Arm velocities, and at velocities slightly lower than the RRL velocity (102.4 \kms, \citet{Lockman89}), but not at +135\kms Expanding Arm velocities; therefore $R_{gal}<3$kpc is assumed.  See Section \ref{HIIdist} for a previous discussion of this \HII region.
\subsection*{\objectname[LPH96 002.819-0.132]{G002.819-00.132}}
The \HI absorption spectrum of G002.819-00.132 does not give conclusive evidence for either a near, nor far, KDAR.
\subsection*{\objectname[GAL 002.90+00.0]{G002.901-00.006}}
The negative RRL velocity suggests a EIG or outer Galaxy location.  Absorption at Near 3kpc Arm velocities infers a $D_{los}>5$kpc; but a lack of absorption associated with any other EIG feature does not allow the EIG/outer Galaxy location degeneracy to be resolved. \citet{Quireza06} give a location in the Outer Galaxy.
\subsection*{G002.961-0.053}
The \HI absorption spectrum is most likely a confusion of multiple \HII regions - there are two RRL velocities (18.1 and -3.5 \kms, GBTHRDS).  No KDAR is given.
\subsection*{\objectname[MSX6C G003.2652-00.1035]{G003.270-00.101}}
Absorption at velocities corresponding to both 3kpc Arms suggests a kinematic location in, or beyond, the Far 3kpc Arm.  The near zero systemic velocity then suggests a location near the Solar circle. \citet{Quireza06} give a line of sight distance to G003.270-00.101  of 14 kpc.
\subsection*{\objectname[L89b 3.342-00.079]{G003.342-00.079}}
Significant absorption is seen at both the Near and Far 3kpc Arms (and at $\sim100$\kms).  Using this absorption as a distance indicator, G003.342-00.079 is given a far KDAR.
\subsection*{G003.439-0.349}
G003.439-0.349 is assumed to be located in the Near 3kpc Arm, due to RRL and maser velocities (GBTHRDS, \citet{Caswell10}), as well as \HI absorption, at velocities expected of the Near 3kpc Arm.
\subsection*{G003.449-0.647}
With \HI absorption at Near 3kpc Arm velocities, $D_{los}>5$kpc is assumed.  As the \HII region has a near zero systemic velocity, and no absorption associated with the Far 3kpc Arm, then $R_{Gal}<3$kpc should also apply.
\subsection*{\objectname[PWN J1754-25551]{G003.655-00.111}}
Absorption at velocities corresponding to the Near 3kpc Arm and a near zero RRL velocity (4.6 \kms, \citet{Lockman89}) suggests $R_{gal}<3$kpc.
\subsection*{G003.928-00.116}
Evidence of absorption is present at velocities corresponding to both the Near and Far 3kpc Arms.  As a result, G003.928-00.116 is given a far side KDAR.  
\subsection*{\objectname[LPH96 00.3949-0.100]{G003.949-00.100}}
The \HI absorption spectrum is of poor quality, typical of the diffuse regions of the \cite{LPH96} catalog.  No KDAR is given, however the small RRL velocity (6.5\kms) suggests a possible EIG location.
\subsection*{G004.346+00.115}
The \HI absorption spectrum of G004.346+00.115 does not give conclusive evidence for either a near, nor far, KDAR.
\subsection*{\objectname[KC97c G004.4+00.1]{G004.412+00.118}}
Absorption is present at velocities corresponding to the Near and Far 3kpc Arms.  This suggests a location of $R_{gal}>3$kpc on the far side of the GC; i.e. a far side KDAR.
\subsection*{G004.527-00.136}
Absorption at Near 3kpc Arm velocities and evidence of absorption at Far 3kpc Arm velocities suggests a far KDAR.
\subsection*{\objectname[KC97c G004.6-00.1]{G004.568-00.118}}
Absorption at velocities associated with the Near 3kpc Arm suggest $D_{los}>5$kpc.
\subsection*{\objectname[GAL 005.19-00.28]{G005.193-00.284}}
Absorption at Near 3kpc and Connecting Arm velocities suggests $D_{los}>7$kpc; that is, the \HII region must be located behind the Connecting Arm along the line of sight.
\subsection*{\objectname[GRS 005.48-00.24]{G005.479-00.241}}
Significant absorption is present at velocities associated with both the Near and Far 3kpc Arms, resulting in a far side KDAR.
\subsection*{\objectname[KKL96 Compact 2]{G005.524+00.033}}
Absorption is present at velocities associated with the Near and Far 3kpc Arms, suggesting a far side KDAR.
\subsection*{\objectname[GAL 005.63+00.23]{G005.633+00.240}}
The \HI absorption spectrum of G005.633+00.240 does not give conclusive evidence for either a near, nor far, KDAR.
\subsection*{G005.889-00.427}
Absorption is not seen towards any EIG features, suggesting a near side KDAR.  \citet{Downes80} also provided a near side KDAR, however, \citet{Quireza06} give a line of sight distance of 14.5 kpc, placing the \HII region on the far side of the GC.
\subsection*{G006.014-00.364}
There are two RRL velocities reported for G006.014-00.364 (14.2 and -31.9 \kms, GBTHRDS) suggesting that there are multiple sources along the line of sight.  No KDAR is given.
\subsection*{\objectname[PMN J1759-2345]{G006.083-00.117}}
Absorption at the $3\sigma_e{^{-\tau}}$ level is seen at velocities associated with the Near and Far 3kpc Arms, suggesting a far side KDAR.  In addition, absorption at $\sim+135$ is present.
\subsection*{\objectname[GAL 006.1-00.6]{G006.148-00.635}}
The \HI absorption spectrum of G006.148-00.635 does not give conclusive evidence for either a near, nor far, KDAR.  However, absorption is present at velocities associated with the Near 3kpc Arm; suggestive of $D_{los}>5$kpc.
\subsection*{G006.160-00.608}
\HI absorption is present at velocities corresponding to the Connecting Arm, but not the Near 3kpc Arm (possibly due to emission fluctuations). 
\subsection*{\objectname[GAL 006.225-00.569]{G006.225-00.569}}
The \HI absorption spectrum of G006.225-00.569 does not give conclusive evidence for either a near, nor far, KDAR.  As with G006.160-00.608, absorption is seen at velocities corresponding with the Connecting Arm, but not the Near 3kpc Arm (which precedes the Connecting Arm along the line of sight).
\subsection*{\objectname[KC97c G006.4-00.5]{G006.398-00.474}}
Absorption is present at velocities corresponding to the Connecting Arm, but not the Near 3kpc Arm (see above).
\subsection*{\objectname[GRS 006.57-00.10]{G006.553-00.095}}
Perhaps the most well behaved absorption spectrum in this work; significant absorption is seen in the Near and Far 3kpc Arms as well as the Connecting Arm, strongly indicative of a far side KDAR.
\subsection*{\objectname[KC97c G006.6-00.3]{G006.565-00.297}}
Again there is no absorption seen at velocities corresponding to the Near 3kpc Arm (see G006.160-00.60, G006.225-00.569 and G006.398-00.474 above), but in this case the lack of absorption is probably due to emission fluctuations.  Absorption at velocities corresponding to the Connecting and Far 3kpc Arms suggests a far side KDAR.
\subsection*{\objectname[LPH96 006.616-0.545]{G006.616-00.545}}
The \HI absorption spectrum of G006.616-00.545 does not give conclusive evidence for either a near, nor far, KDAR.
\subsection*{G007.041+00.176}
Significant absorption is present at velocities corresponding to the Near and Far 3kpc Arms, a far side KDAR is given.
\subsection*{G007.176+00.086}
The \HI absorption spectrum of G006.616-00.545 does not give conclusive evidence for either a near, nor far, KDAR.
\subsection*{G007.254-00.073}
G007.254-00.073 has two known RRL velocities (47 and 17.5 \kms, GBTHRDS), suggesting multiple sources along the line of sight. 
\subsection*{G007.266+00.186}
The \HI absorption spectrum of G007.266+00.186 demonstrates absorption at velocities associated with the Near 3kpc and Connecting Arms, which suggests a location $D_{los}>7$kpc.  In addition, a near zero RRL velocity (-4.4 \kms, GBTHRDS), and a lack of absorption at Far 3kpc Arm velocities suggests $R_{gal}<3$kpc.
\subsection*{G007.299-00.116}
The \HI absorption spectrum of G007.299-00.116 does not give conclusive evidence for either a near, nor far, KDAR.
\subsection*{\objectname[GPSR 007.420+0.366]{G007.420+00.366}}
The \HI absorption spectrum of G007.420+00.366 shows absorption at velocities associated with the Near and Far 3kpc Arms,as a result the \HII region is given a far side KDAR.
\subsection*{G007.466-00.279}
Absorption is present at velocities associated with the Near 3kpc Arm, but no further evidence for a KDAR is forthcoming from the \HI absorption spectrum.
\subsection*{\objectname[GAL 007.47+00.06]{G007.472+00.060}}
Significant absorption at velocities corresponding to both the Near and Far 3kpc Arms strongly suggests a far side KDAR, however the RRL velocity (-17.8\kms, \citet{Lockman89}) is indicative of a location in the Near 3kpc Arm.  The \HII region also presents significant absorption at $\sim+135$\kms (see G006.083-00.117, above).
\subsection*{G007.700-00.079}
The \HII region has two observed RRL velocities, one of which is associated with the velocity expected of the Connecting Arm.  No KDAR is given.
\subsection*{\objectname[LPH96 007.768+0.014]{G007.768+00.014}}
The \HI absorption spectrum of G007.768+0.014 does not give conclusive evidence for either a near, nor far, KDAR.
\subsection*{G007.806-00.621}
Evidence of absorption at velocities corresponding to the Near 3kpc Arm, Connecting Arm implies a distance along the line of sight as least as far as the Connecting Arm.
\subsection*{G008.005-00.484}
Absorption is present at velocities consistent with the Near 3kpc and Connecting Arms, but with no other EIG features.  The lower line of sight distance limit is therefore $D_{los}>7$kpc.  
\subsection*{\objectname[KC97c G008.0-00.2]{G008.006-00.156}}
Evidence of absorption at velocities corresponding to the Near 3kpc Arm, Connecting Arm and Far 3kpc Arm strongly implies a far side KDAR.
\subsection*{\objectname[LPH96 008.094+0.085]{G008.094+00.085}}
The \HI absorption spectrum of G008.094+00.085 does not give conclusive evidence for either a near, nor far, KDAR.
\subsection*{G008.103+00.340}
The \HI absorption spectrum of G008.103+00.340 does not give conclusive evidence for either a near, nor far, KDAR.  However absorption is present at velocities corresponding to the Near 3kpc Arm.
\subsection*{\objectname[PMN J1803-2148]{G008.137+00.228}}
Absorption is not seen towards any EIG features, suggesting a near side KDAR in agreement with \citet{WAM82} and \citet{Quireza06}.
\subsection*{\objectname[KW90 C]{G008.362-00.303}}
The \HI absorption spectrum of G008.362-00.303 does not give conclusive evidence for either a near, nor far, KDAR.  However absorption is present at velocities corresponding to the Near 3kpc Arm.
\subsection*{G008.373-00.352}
Absorption is present at velocities consistent with the Near 3kpc and Connecting Arms, but with no other EIG features. 
\subsection*{G008.432-00.276}
Evidence of absorption due to both the Near and Far 3kpc Arms suggests a far side KDAR
\subsection*{\objectname[KC97c G008.7-00.4]{G008.666-00.351}}
 \citet{maserKDA} and \citet{Downes80} both give a near side KDAR for the \HII region.  A near side KDAR is also given by this work - note that at this longitude the expected velocity range of the Near 3kpc Arm overlaps the expected velocities of general circular rotation.
\subsection*{G008.830-00.715}
Absorption is not seen towards any EIG features, suggesting a near side KDAR.
\subsection*{\objectname[KC97c G008.9-00.3]{G008.865-00.323}}
The \HI absorption spectrum of G008.362-00.303 does not give conclusive evidence for either a near, nor far, KDAR.
\subsection*{G009.178+00.043}
There \HI absorption spectrum suffers from emission fluctuations, the \HII region must be located at least as far as the Near 3kpc Arm along the line of sight.
\subsection*{\objectname[GAL 009.62+00.19]{G009.615+00.198}}
Kinematic distance analyses can be greatly affected by velocity crowding and a decrease of cold, dense \HI in the EIG.  For example, \citet{Sanna09} thoroughly investigated the high mass star formation region G9.62+0.20 - comprised of several \HII regions - and find from trigonometric parallax that it has a distance of $5.2\pm0.6$ kpc, placing it in the 3 kpc Arm.  This distance is at odds with the kinematically determined distances for the region (0.36 and 16.4 kpc, based on the systemic velocity of the region, 4.1 \kms).  Inspection of the \HI absorption spectrum of G009.615+00.198 rules out the far kinematic distance as there is no significant absorption at the velocities of far side EIG features (i.e. the Far 3kpc Arm).  
\subsection*{G009.925-00.745, G009.682+00.206, \objectname[KC97c G009.7-00.8]{G009.717-00.832}, G009.741+00.842, \objectname[WHR97 18073-2046]{G009.875-00.749} and G009.982-00.752}
At this longitude the velocities associated with the Near 3kpc Arm and normal circular rotation overlap.  No KDARs are given.


\begin{thebibliography}{}
\bibitem[Anderson et al.(2012)]{Anderson12} Anderson, L.~D., Bania, T.~M., Balser, D.~S., \& Rood, R.~T.\ 2012, \apj, 754, 62
\bibitem[Anderson et al.(2011)]{GBTHRDS}Anderson, L.~D., Bania, T.~M., Balser, D.~S., \& Rood, R.~T.\ 2011, \apjs, 194, 32
\bibitem[Anderson \& Bania(2009)]{AndersonBania09} Anderson, L.~D. \& Bania, T.~M.\ 2009, \apj, 690, 706
\bibitem[Baba et al.(2010)]{BSW10} Baba, J., Saitoh, T.~R., \& Wada, K.\ 2010, \pasj, 62, 1413
\bibitem[Bania(1980)]{Bania80} Bania, T.~M.\ 1980, \apj, 242, 95
\bibitem[Bania \& Lockman(1984)]{BaniaLockman84} Bania, T.~M., \& Lockman, F.~J.\ 1984, \apjs, 54, 513
\bibitem[Bania et al.(1986)]{Bania86} Bania, T.~M., Stark, A.~A., \& Heiligman, G.~M.\ 1986, \apj, 307, 350
\bibitem[Bania et al.(2010)]{BaniaAnderson10} Bania, T.~M., Anderson, L.~D., Balser, D.~S., \& Rood, R.~T.\ 2010, \apjl, 718, L106
\bibitem[Bania et al.(2012)]{Bania12} Bania, T.~M., Anderson, L.~D., \& Balser, D.~S.\ 2012, \apj, 759, 96 
\bibitem[Blitz et al.(1993)]{Blitz93} Blitz, L., Binney, J., Lo, K.~Y., Bally, J., \& Ho, P.~T.~P.\ 1993, \nat, 361, 417
\bibitem[Brand \& Blitz(1993)]{BB93} Brand, J., \& Blitz, L.\ 1993, \aap, 275, 67 
\bibitem[Burton \& Liszt(1983)]{BurtonLiszt83} Burton, W.~B., Liszt, H.~S.\ 1983, \aaps, 52, 63
\bibitem[Caswell \& Haynes(1982)]{CH82} Caswell, J.~L., \& Haynes, R.~F.\ 1982, \apjl, 254, L31
\bibitem[Caswell \& Haynes(1987)]{CH87} Caswell, J.~L., \& Haynes, R.~F.\ 1987, \aap, 171, 261
\bibitem[Caswell et al.(2010)]{Caswell10} Caswell, J.~L., Fuller, G.~A., Green, J.~A., et al.\ 2010, \mnras, 404, 1029 
\bibitem[Churchwell et al.(2009)]{GLIMPSE} Churchwell, E., Babler, B.~L., Meade, M.~R., et al.\ 2009, \pasp, 121, 213
\bibitem[Clemens(1985)]{Clemens85} Clemens, D.~P.\ 1985, \apj, 295, 422 
\bibitem[Condon et al.(1998)]{NVSS} Condon, J.~J., Cotton, W.~D., Greisen, E.~W., et al.\ 1998, \aj, 115, 1693
\bibitem[Dame et al.(2001)]{DameCO} Dame, T.~M., Hartmann, D., \& Thaddeus, P.\ 2001, \apj, 547, 792 
\bibitem[Dame \& Thaddeus(2008)]{DameThaddeus08} Dame, T.~M., \& Thaddeus, P.\ 2008, \apjl, 683, L143
\bibitem[Dame \& Thaddeus(2011)]{DameThaddeus11} Dame, T.~M., \& Thaddeus, P.\ 2011, \apjl, 734, L24 
\bibitem[Dickey et al.(2003)]{Dickey03} Dickey, J.~M., McClure-Griffiths, N.~M., Gaensler, B.~M., \& Green, A.~J.\ 2003, \apj, 585, 801 
\bibitem[Dickey \& Lockman(1990)]{DickeyLockman90} Dickey, J.~M., \& Lockman, F.~J.\ 1990, \araa, 28, 215 
\bibitem[Downes et al.(1980)] {Downes80} Downes, D., Wilson, T.~L., Bieging, J., \& Wink, J.\ 1980, \aaps, 40, 379 
\bibitem[Fux(1999)]{Fux99} Fux, R.\ 1999, \aap, 345, 787
\bibitem[Green et al.(2009)]{Green09} Green, J.~A., McClure-Griffiths, N.~M., Caswell, J.~L., et al.\ 2009, \apjl, 696, L156
\bibitem[Green \& McClure-Griffiths(2011)]{maserKDA} Green, J.~A., \& McClure-Griffiths, N.~M.\ 2011, \mnras, 417, 2500 
\bibitem[Green et al.(2011)]{GreenMasers} Green, J.~A., Caswell, J.~L., McClure-Griffiths, N.~M., et al.\ 2011, \apj, 733, 27 
\bibitem[Gooch(1996)]{KARMA} Gooch, R.\ 1996, Astronomical Data Analysis Software and Systems V, 101, 80
\bibitem[Hachisuka et al.(2009)]{Hachisuka09} Hachisuka, K., Brunthaler, A., Menten, K.~M., et al.\ 2009, \apj, 696, 1981 
\bibitem[Honma et al.(2007)]{Honma07} Honma, M., Bushimata, T., Choi, Y.~K., et al.\ 2007, \pasj, 59, 889 
\bibitem[Jones \& Dickey(2012)]{Jones12} Jones,C., \& Dickey, J.\ 2012, \apj, 753, 62
\bibitem[Kolpak et al.(2003)]{Kolpak03} Kolpak, M.~A., Jackson, J.~M., Bania, T.~M., Clemens, D.~P., \& Dickey, J.~M.\ 2003, \apj, 582, 756 
\bibitem[Lang et al.(2010)]{Lang10} Lang, C.~C., Goss, W.~M., Cyganowski, C., \& Clubb, K.~I.\ 2010, \apjs, 191, 275
\bibitem[Liszt(2008)]{Liszt08} Liszt, H.~S.,\ 2008, \aap, 486, 467
\bibitem[Liszt \& Burton(1980)]{LisztBurton80} Liszt, H.~S, Burton, W.~B. \ 1980, \apj, 236, 779
\bibitem[Lockman(1989)]{Lockman89} Lockman, F.~J.\ 1989, \apjs, 71, 469
\bibitem[Lockman et al.(1996)]{LPH96} Lockman, F.~J., Pisano, D.~J., \& Howard, G.~J.\ 1996, \apj, 472, 173
\bibitem[Marshall et al.(2008)]{Marshall08} Marshall, D.~J., Fux, R., Robin, A.~C., \& Reyl{\'e}, C.\ 2008, \aap, 477, L21
\bibitem[McClure-Griffiths et al.(2005)]{SGPS} McClure-Griffiths, N.~M., Dickey, J.~M., Gaensler, B.~M., et al.\ 2005, \apjs, 158, 178 
\bibitem[McClure-Griffiths \& Dickey(2007)]{MCGD07} McClure-Griffiths, N.~M., \& Dickey, J.~M.\ 2007, \apj, 671, 427 
\bibitem[McClure-Griffiths et al.(2012)]{HIGC} McClure-Griffiths, N.~M., Dickey, J.~M., Gaensler, B.~M., et al.\ 2012, \apjs, 199, 12 
\bibitem[Mois{\'e}s et al.(2011)]{Moises11} Mois{\'e}s, A.~P., Damineli, A., Figuer{\^e}do, E., et al.\ 2011, \mnras, 411, 705 
\bibitem[Morris \& Serabyn(1996)]{MorrisSerabyn96} Morris, M., \& Serabyn, E.\ 1996, \araa, 34, 645 
\bibitem[Oort(1977)]{Oort77}Oort, J.~H.\ 1977, \araa, 15, 295
\bibitem[Paladini et al.(2003)]{Pal03} Paladini, R., Burigana, C., Davies, R.~D., et al.\ 2003, \aap, 397, 213 
\bibitem[Quireza et al.(2006)]{Quireza06} Quireza, C., Rood, R.~T., Bania, T.~M., Balser, D.~S., \& Maciel, W.~J.\ 2006, \apj, 653, 1226 
\bibitem[Reid et al.(2009)]{Reid09} Reid, M.~J., Menten, K.~M., Zheng, X.~W., et al.\ 2009, \apj, 700, 137 
\bibitem[Rodr{\'{\i}}guez-Fern{\'a}ndez(2006)]{RF06} Rodr{\'{\i}}guez-Fern{\'a}ndez, N.~J.\ 2006, Journal of Physics Conference Series, 54, 35 
\bibitem[Rodriguez-Fernandez et al.(2006)]{RFetal06} Rodriguez-Fernandez, N.~J., Combes, F., Martin-Pintado, J., Wilson, T.~L., \& Apponi, A.\ 2006, \aap, 455, 963 
\bibitem[Rodr{\'{\i}}guez-Fern{\'a}ndez(2011)]{RF11} Rodr{\'{\i}}guez-Fern{\'a}ndez, N.~J.\ 2011, Memorie della Societa Astronomica ItalianaSupplementi, 18, 195
\bibitem[Roy(2003)]{Roy03} Roy, S.\ 2003, \aap, 403, 917 
\bibitem[Sanna et al.(2009)]{Sanna09} Sanna, A., Reid, M.~J., Moscadelli, L., et al.\ 2009, \apj, 706, 464
\bibitem[Sewilo et al.(2004)]{Sewilo04} Sewilo, M., Watson, C., Araya, E., et al.\ 2004, \apjs, 154, 553
\bibitem[Simonson \& Madder(1973)]{SM73} Simonson, S.~C., \& Madder, G.~L., \ 1973, \aap, 27, 337 
\bibitem[Uchida et al.(1992)]{Uchida92} Uchida, K., Morris, M., \& Yusef-Zadeh, F.\ 1992, \aj, 104, 1533 
\bibitem[Urquhart et al.(2012)]{Urquhart12} Urquhart, J.~S., Hoare, M.~G., Lumsden, S.~L., et al.\ 2012, \mnras, 420, 1656
\bibitem[van Woerden et al.(1957)]{vanWoerden57} van Woerden, H., Rougoor, G.~W., \& Oort, J.~H.\ 1957, Academie des Sciences Paris Comptes Rendus, 244, 1691 
\bibitem[Wink et al.(1982)]{WAM82} Wink, J.~E., Altenhoff, W.~J., \& Mezger, P.~G.\ 1982, \aap, 108, 227

\end{thebibliography}
\end{document}